\documentclass[prd,nofootinbib,floatfix,eqsecnum,superscriptaddress]{revtex4-2}
\usepackage[utf8]{inputenc}
\usepackage{amsmath}
\usepackage{graphicx}
\usepackage{subfig}
\usepackage{color}
\graphicspath{ {images/} }
\def\apjl{Astrophysical\ Journal,\ Letters}
\def\aap{Astron.\ Astrophys.}
\def\mnras{Mon.\ Notices Royal Astron.\ Soc.}
\def\jcap{J.\ Cosmol.\ Astropart.\ Phys.}
\newcommand{\change}[1]{\textcolor{black}{#1}}

\begin{document}

\title{Optimal method for reconstructing polychromatic maps from broadband observations with an asymmetric antenna pattern}
\author{Brianna Cantrall}
\affiliation{Department of Physics, University of Richmond}
\author{Solomon Quinn}
\affiliation{Committee on Computational and Applied Mathematics, University of Chicago}
\author{Emory F. Bunn}
\affiliation{Department of Physics, University of Richmond}

\begin{abstract}
    Broadband time-ordered data obtained from telescopes with a wavelength-dependent, asymmetric beam pattern can be used to extract maps at multiple wavelengths from a single scan. This technique is especially useful when collecting data on cosmic phenomena such as the Cosmic Microwave Background (CMB) radiation, as it provides the ability to separate the CMB signal from foreground contaminants. We develop a method to determine the optimal linear combinations of wavelengths (``colors'') that can be reconstructed for a given telescope design and the number of colors that are measurable with high signal-to-noise ratio. The optimal colors are found as eigenvectors of a matrix derived from the inverse noise covariance matrix. When the telescope is able to scan the sky isotropically, it is useful to transform to a spherical harmonic basis, in which this matrix has a particularly simple form. We propose using the optimal colors determined from the isotropic case even when the actual scanning pattern is not isotropic (e.g., covers only part of the sky). We perform simulations showing that maps in multiple colors can be reconstructed accurately from both full-sky and partial-sky scans. Although the original motivation for this research comes from mapping the CMB, this method of polychromatic map-making will have broader applications throughout astrophysics.
\end{abstract}

\maketitle

\section{\label{introduction}Introduction} 
   
In many if not most astrophysical observations, both spatial and spectral information about the incoming radiation are invaluable. Unfortunately, some astrophysical surveys are made with broadband detectors that provide little spectral information. This problem is particularly noteworthy in observations of the cosmic microwave background (CMB) radiation, for which more fine-grained spectral information would be invaluable in separating the CMB signal from foreground contamination. 

Reliably separating the different physical components contributing to CMB maps is an extremely important step in drawing conclusions from these observations (e.g., \cite{planckcomponentseparation} and references therein). These inferences must generally be drawn from a relatively small number of broadband maps. Numerous methods have been developed to achieve this goal (e.g., \cite{delabrouillecardoso,DCP,remazeilles,roma}). To do this separation as well as possible, one must extract every bit of information from the data one has. In this paper, we examine the prospect of extracting spectral information from a single broadband total-intensity survey.

When a telescope scans the sky with a broadband detector, each observation in the time-ordered data is a weighted average of signals at different spatial points and at different wavelengths. Because the antenna pattern is inevitably wavelength-dependent, it is in principle possible to reconstruct spectral information even though each individual observation lacks spectral resolution \cite{stolpovskiy,malu,qubic1,qubic2,quinnaas,mooneyaas}. Because such spectral information can be extremely valuable, we develop in this paper a systematic approach to determining the amount and kind of information that can be reconstructed.

Reconstruction of spectral information requires a telescope whose antenna pattern lacks azimuthal symmetry, and which scans the sky in a way that causes it to ``hit" a given point in different orientations. We illustrate this with a toy model. Imagine that a patch of sky contains a wavelength-dependent sinusoidal variation $I(x,\lambda) = f(\lambda)\sin(kx)$, where $(x,y)$ denote position on the sky. A telescope observes this patch with an antenna pattern that smooths the sky in the $x$ direction but not in the $y$ direction. If the smoothing is wavelength-dependent, then different wavelengths will be suppressed by different amounts, and the observed signal will be a sine wave whose  amplitude is a weighted sum of the amplitudes at different wavelengths, $\int f(\lambda)w(\lambda)d\lambda$, for some weighting function $w(\lambda)$ that accounts for the wavelength-dependent smoothing.
Now suppose that the telescope is rotated so that it smooths in the $y$ direction but not the $x$ direction. Then the observed signal amplitude will be an unweighted sum proportional to $\int f(\lambda)d\lambda$. Comparing these two signals would allow the observer to reconstruct a ``color" in addition to the overall amplitude of the signal, even though each individual observation is a broadband total-power measurement.

In the CMB context, the instrument for which this observation is most likely to be relevant is the Q \& U Bolometric Interferometer for Cosmology 
(QUBIC) (\cite{qubic1} and references therein), because the technique of bolometric interferometry combines highly asymmetric beam patterns with broadband detectors. However, the technique may be applicable in other contexts involving telescopes with highly asymmetric beams, such as the CHIME 21-cm experiment \cite{chime}, which uses long parabolic antennas.

In this paper, we do not focus on application to a particular telescope, but on developing a formalism for quantifying the available information for any given instrument. We will present illustrative examples, some of which are loosely inspired by QUBIC and by a long parabolic dish, but they are not intended to mimic any particular experiment in detail.

Our approach is based on a generalization of the well-known minimum-variance mapmaking procedure for turning time-ordered data into sky maps (e.g., \cite{Tegmark}). In the original formalism, one reconstructs a single spatially-dependent signal $\vec s$, which is an $N_p$-dimensional vector giving the signal in each of $N_p$ pixels. To extend this formalism to the polychromatic case, we divide the observed wavelength range into $N_f$ bins and consider a wavelength-dependent signal at each pixel. Our signal vector thus has dimension $N_pN_f$.

It is straightforward to write down the formal minimum-variance reconstruction of this signal vector from a given vector $\vec d$ of time-ordered data. Because the wavelength-dependent information is limited, this reconstruction has extremely large error bars and is of limited use. Certain linear combinations of wavelengths, however, can be reliably reconstructed. That is, we can choose weights $v_1,\ldots,v_{N_f}$ and reconstruct a signal at each pixel $p$ of the form
\begin{equation}
u_p = \sum_f v_fs_{pf},
\end{equation}
where $s_{pf}$ refers to the signal at pixel $p$ and wavelength $f$.

We will identify the weights $\vec v$ that can be reconstructed most accurately. The noise level in these reconstructions can also be computed, so we can determine how many such wavelength combinations are worth reconstructing in any given experiment. 

The remainder of this paper is organized as follows. Section \ref{method} describes the mathematical and computational formalism. Section \ref{sec:results} presents illustrative results of tests performed on simulated data. A discussion of some conclusions is found in Section \ref{sec:conclusion}. Some mathematical details are relegated to appendices.


\section{\label{method}Method} 
    This section contains information about the mathematical methods behind this research. We begin by reviewing the map-making problem in the monochromatic case  (e.g., \cite{Tegmark,tegmark-designer,madcap}), and then extend those results to a case containing polychromatic data. We then describe the method to be used when mapping signals obtained from such data, as well as various operations to perform to optimize these calculations. Finally, we show that there are specific modes that are filtered out of the reconstructed maps; the same modes must be filtered out of the true signal when comparing the two.


    \subsection{\label{sec:mono}Monochromatic Map-Making} 
    The data collected by a telescope can be modeled by the linear equation
    \begin{equation}
        \vec{d} = \mathbf{A}\vec{s} + \vec{n},
    \end{equation}
    where the data vector $\vec{d}$ of size $N_t$ contains all of the time-ordered data (TOD) gathered by a given telescope, the pointing matrix $\mathbf{A}$ of size $N_t \times N_p$ contains information about which pixels each TOD element is sensitive to, the signal vector $\vec{s}$ of size $N_p$ contains the intensity of light as a function of position (i.e., pixel), and the vector $\vec{n}$ of size $N_t$ contains instrument noise. The quantities $N_t$ and $N_p$ are the number of timesteps and the number of pixels, respectively. 
    
   The elements of the pointing matrix can
   be written as
    \begin{equation} \label{eq:antenna}
        A_{tp} = \mathcal{A}(\mathbf{R}_t(\hat{r}_p))A_{\mathrm{pix}},
    \end{equation}
    where $\mathcal{A}(\hat{r})$ is the antenna pattern at some specific reference orientation, $A_{\mathrm{pix}}$ is the area of a pixel, introduced here for later convenience in Section \ref{sec:SH} when transforming to a spherical harmonic basis, and $\mathbf{R}_t$ is a rotation matrix that transforms $\mathcal{A}(\hat{r})$ to the orientation at time $t$ and pixel $p$. 
    
    It is often the case that the antenna pattern is composed of a superposition of many individual peaks, each of which having the same Gaussian-like shape. For example, in a single-difference measurement, $\mathcal{A}$ would consist of two offset antenna patterns, one with negative weight. Of more relevance to the current work is the case of interferometer-like antenna patterns such as that produced by QUBIC, which consists of many identically-shaped peaks.
    
    In this case, it is numerically preferable to replace the true sky signal $\vec s$ with a smoothed signal that has been convolved with the shape of a single peak in $\mathcal{A}$ \cite{tegmark-designer}. After making this replacement, the pointing matrix $\mathbf{A}$ becomes very sparse, with each row containing only a few nonzero elements corresponding to the $\delta$-function-like peaks at the individual pixels corresponding to the peaks of $\mathcal{A}$. This renders computations and storage of $\mathbf{A}$ much more efficient.

    With this substitution, the map-making algorithm returns an estimate of the smoothed signal, not the true signal. If one wished (unwisely) to try to reconstruct features smaller than the beam scale, one could attempt to deconvolve the reconstructed map.

    When solving for the best estimate of the signal, we want to retrieve all information possible about the signal given the data. However, the pointing matrix is not square and often very large, making inversion impossible. There are many methods of finding a solution to this problem, and in this research we use the minimum-variance reconstruction method used by the COBE-DMR team \cite{Janssen1992} (among others). The estimated signal map is
    \begin{equation} \label{eq:s}
        \hat{s} = (\mathbf{A}^T\mathbf{N}^{-1}\mathbf{A})^{-1}\mathbf{A}^T\mathbf{N}^{-1}\vec{d},
    \end{equation}
    where $\hat{s}$ is a vector of size $N_p$ that is the best estimate of $\vec{s}$, and $\mathbf{N}$ is the noise covariance matrix of size $N_t \times N_t$, such that $\mathbf{N}=\langle \vec{n}\vec{n}^{\,T} \rangle$. The resulting $\hat{s}$ is the unbiased, maximum-likelihood estimator of $\vec{s}$, meaning that, by the Cram\'er-Rao inequality, it is also the minimum-variance unbiased estimator. The noise covariance matrix of $\hat s$ is
    \begin{equation} \label{eq:M}
        \mathbf{M}=(\mathbf{A}^T\mathbf{N}^{-1}\mathbf{A})^{-1}
    \end{equation} 
    This method of map-making loses no information contained in the TOD \cite{Tegmark}.
    
    
    \subsection{\label{sec:poly}Polychromatic Map-Making}
    We now consider maps that are functions of wavelength. To be specific, we discretize the wavelength range under consideration into $N_f$ narrow wavebands, which we call wavelength bins, each with a different signal map. The signal vector $\vec s$ consists of all of these maps stacked on top of each other and thus has size 
    $N_pN_f$.
    We will discuss the required number of wavelength bins in Section \ref{sec:results}. 
    
    The pointing matrix $\mathbf{A}$ now has dimensions $N_t\times N_pN_f$. The elements of $\mathbf{A}$ are
    \begin{equation} \label{eq:polyA}
        A_{t(pf)} = \mathcal{A}(\mathbf{R}_t(\hat{r}_p);\lambda_f)A_{\mathrm{pix}}.
    \end{equation}
    This expression differs from equation (\ref{eq:antenna}) in that the columns of $\mathbf{A}$ are now labeled by a pair of indices $(pf)$ corresponding to pixel and wavelength bin, and the antenna pattern now depends on the wavelength $\lambda_f$.
    Continuing to treat $\mathbf{A}$ as a sparse matrix as described in Section \ref{sec:mono}, each row of $\mathbf{A}$ now contains a copy of the antenna pattern for each waveband. We compute the pointing matrix for each antenna pattern separately as sparse matrices at a given waveband, and then horizontally stack these matrices, creating a matrix composed of $N_f$ sub-matrices. 
    
    One can perform the minimum-variance map reconstruction (\ref{eq:s}) and reconstruct the polychromatic signal with noise covariance $\mathbf{M}$ given by equation (\ref{eq:M}). However, the $N_pN_f\times N_pN_f$ covariance matrix is unwieldy. Moreover, in practice the inverse noise covariance matrix
       \begin{equation}
        \mathbf{M}^{-1} = \mathbf{A}^T \mathbf{N}^{-1} \mathbf{A}
    \end{equation}
is typically nearly singular, so the inversion to produce $\mathbf{M}$ is unstable. Even when the inversion can be performed accurately, the diagonal elements (i.e., the noise variances in the reconstructed signal) are so large that the reconstructed signal is not particularly useful. This occurs because there is simply not enough information to reconstruct the full wavelength-dependent signal with reasonable accuracy.

Rather than reconstructing the entire polychromatic signal, we will reconstruct only certain linear combinations of wavelengths chosen to have low noise. To be specific, we will find a set of $N_f$-dimensional vectors $\vec v^{(1)},\vec v^{(2)},\ldots$ representing different combinations of wavelengths that we wish to estimate. Each vector $\vec v^{(i)}$ corresponds to a signal map $\vec  u^{(i)}$, given by
    \begin{equation}
    \label{eq:u_map}
        u^{(i)}_p = \sum_f v^{(i)}_f s_{pf}.
    \end{equation}
For instance, if $v^{(i)}_f$ is the same for all $f$, the resulting map $\vec u^{(i)}$ would be the total power across all wavelengths, while a $\vec v^{(i)}$ that was positive on one half of the wavelength range and negative on the other would give a ``color" map.

Suppose that we have chosen an orthonormal basis of such $N_f$ vectors. Then there is an orthogonal operator $\mathbf{U}$ such that
\begin{equation}
    \vec u = \mathbf{U}\vec s,
\end{equation}
where $\vec u$ is the concatenation of the maps $\vec u^{(i)}$ just as $\vec s$ is the concatenation of the individual wavelength maps. To be specific, $\mathbf{U}$ consists of one copy for every pixel of the orthogonal $N_f\times N_f$ matrix whose rows are the vectors $\vec v^{(i)}$: $U_{(pi)(p'f)} = v_f^{(i)}\delta_{pp'}$.
Our goal will be to choose the vectors $\vec v^{(i)}$ so that the useful information in the data is concentrated in the first few vectors, and we can safely ignore the later, noise-dominated ones.

The optimal estimator of $\vec u$ is easily shown to be 
\begin{equation}
    \hat u = \mathbf{U}\hat s
\end{equation}
as one would expect. That is, each individual ``color'' map has pixel values
       \begin{equation}
        \hat u^{(i)}_p = \sum_f v^{(i)}_f \hat s_{pf}.
    \end{equation}
The estimated map corresponding to the $i$th wavelength combination has inverse noise covariance matrix elements
    \begin{equation} \label{eq:variance}
({M^{(i)}}^{-1})_{pp'}=
\sum_{f,f'}{v}^{(i)}_fM^{-1}_{(pf)(p'f')}{v}^{(i)}_{f'}.
    \end{equation}
    Let $w^{(i)}$ be the trace of this matrix. This quantity is the sum of the inverse noise variances of all pixels in the $i$th map and is a natural figure of merit for the map. It can be written as
    \begin{equation}
        w^{(i)}=\sum_{f,f'}{v}^{(i)}_f\bar{M}^{-1}_{ff'}{v}^{(i)}_{f'},
    \end{equation}
    where the $N_f\times N_f$ matrix $\mathbf{\bar{M}}^{-1}$ has elements 
    \begin{equation}\label{eq:mbarinverse}
   \bar{M}^{-1}_{ff'}=\sum_{p}M^{-1}_{(pf)(pf')}.
    \end{equation}
    
The best linear combination of wavelengths is the one that maximizes the corresponding $w$. This is the eigenvector of $\mathbf{\bar{M}}^{-1}$ with the largest eigenvalue. Indeed, for any number of maps $k$ between 1 and $N_f$, the best $k$ maps to reconstruct are the ones corresponding to the greatest $k$ eigenvectors of $\bar{\mathbf{M}}^{-1}$. See Appendix \ref{eigenvector_proof} for the proof of this claim.

    
    \subsection{Transformation to Spherical Harmonic Space} \label{sec:SH}
    Computing $\mathbf{\bar{M}}^{-1}$ in pixel space remains a computationally expensive task. However, if we assume we are performing an isotropic experiment, which we define as one in which a given telescope is able to scan the whole sky uniformly and in all orientations, we can instead perform this calculation in a basis of spherical harmonic coefficients. In this basis, the noise covariance matrix  is
    \begin{equation}
        \mathbf{\tilde{M}} = (\mathbf{Y}^\dagger \mathbf{A}^\dagger \mathbf{N}^{-1} \mathbf{A} \mathbf{Y})^{-1}, 
    \end{equation}
    where $\mathbf{Y}$ is the linear operator that transforms a set of spherical harmonic coefficients into the corresponding map, whose elements are
    \begin{equation}
        Y_{(pf)(lm)} = Y_{lm}(\hat{r}_p),
    \end{equation}
    independent of $f$. $\mathbf{\tilde{M}}$ is of size $N_pN_f\times N_L$, where $N_L$ is the number of $a_{lm}$ coefficients we are estimating. As shown in Appendix \ref{sec:c_list}, under the assumptions of an isotropic experiment, the matrix is  block diagonal, with no correlation between $lm$ pairs. Moreover, the blocks for all $m$ corresponding to a given $l$ are identical. 
    
    Let $\mathbf{C}_{l}$ be one such block of $\mathbf{\tilde{M}}^{-1}$. Then, holding an $lm$ pair fixed and suppressing the index $l$, the elements of $\mathbf{C}_l$ are
    \begin{equation}
        C_{ff'} = \tilde{M}^{-1}_{(lmf)(lmf')}
        = \sigma^{-2}_t (Y^\dagger A^\dagger A Y)_{ff'},
    \end{equation}
    assuming white noise with $\mathbf{N}=\sigma_t^2\mathbf{1}$. 
    
    As shown in Appendix \ref{sec:c_list}, 
    \begin{equation}
        C_{ff'} = \frac{N_t}{\sigma^2_t}\frac{1}{2l+1} \sum_{m'}A_{lm'f}A^*_{lm'f'},
    \end{equation}
    where $A_{lmf}$ is a coefficient in the spherical harmonic expansion of the antenna pattern at wavelength $\lambda_f$. 
    
    For any given $l$, the eigenvectors of $\mathbf{C}_l$ with largest eigenvalues give the best linear combinations of wavelengths, with inverse noise variances given by the eigenvalues. To obtain the best linear combinations over all $l$, we should ``trace over'' $lm$, just as we trace over all pixels in defining the matrix $\bar{\mathbf{M}}^{-1}$ in equation (\ref{eq:mbarinverse}). In fact, because of the orthonormality in spherical harmonics, this trace, namely
    \begin{equation}
        \sum_l (2l+1)\mathbf{C}_l,
    \end{equation}
    is equivalent to $\bar{\mathbf{M}}^{-1}$ for isotropic experiments (i.e., in the limit $N_t\to\infty$ with all antenna pattern orientations equally sampled).

    Because the spherical-harmonic formalism is extremely efficient, we use it to determine the best linear combinations of wavelengths to reconstruct, even when the actual data do not scan the whole sky isotropically.
    

    \begin{figure*}
        \centering
        \includegraphics[width=0.3\textwidth]{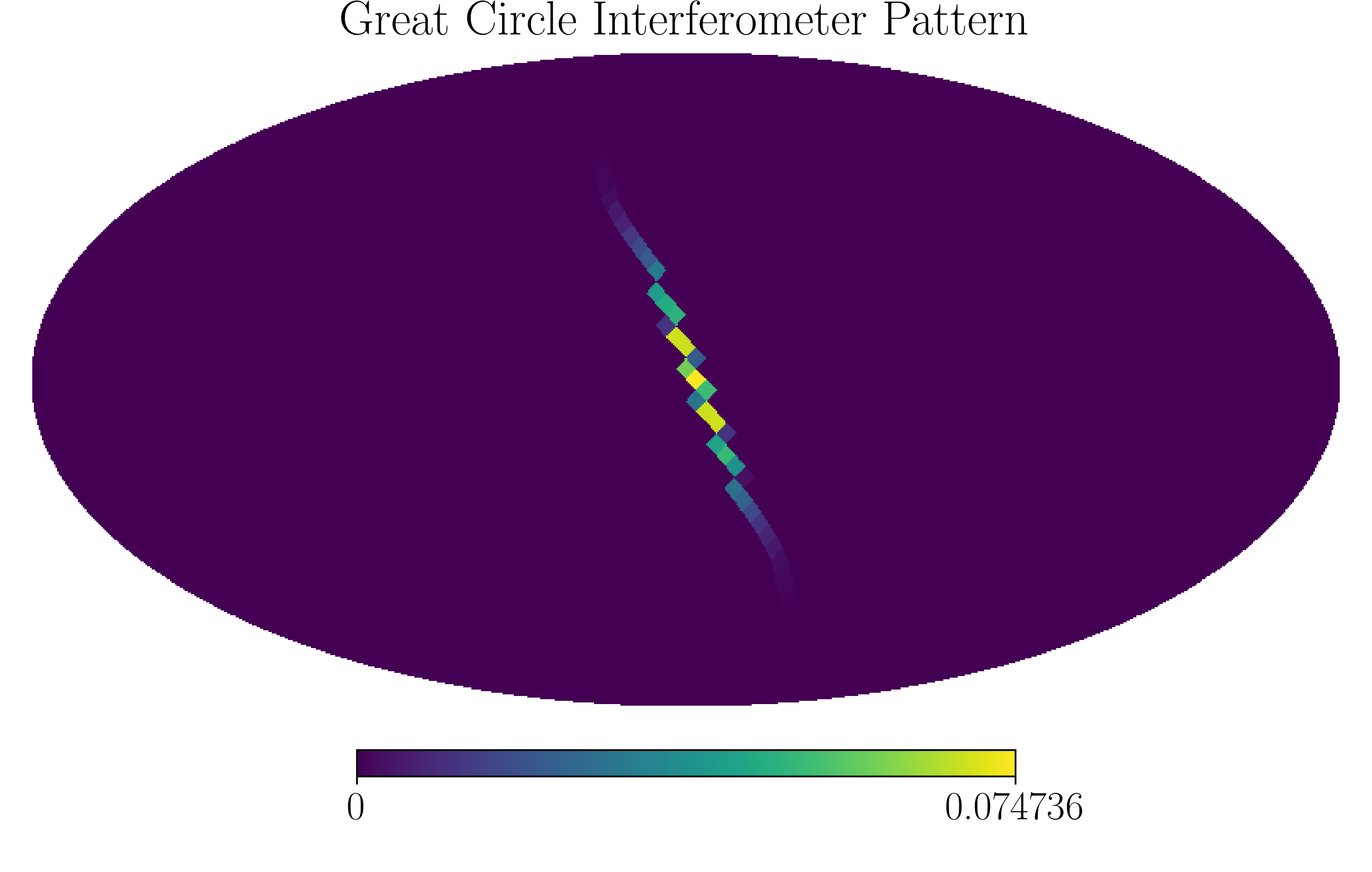}
        \includegraphics[width=0.3\textwidth]{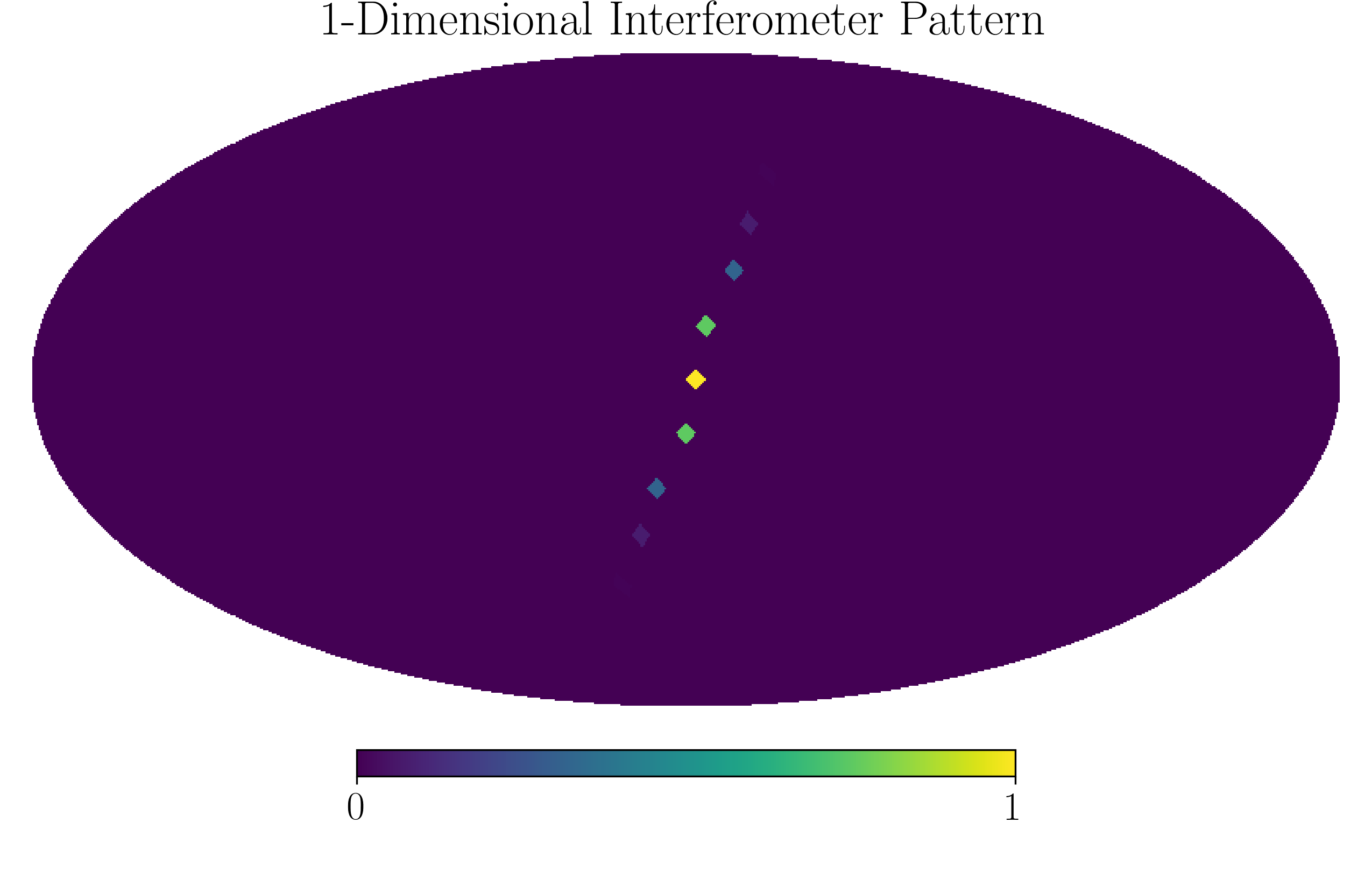}
        \includegraphics[width=0.3\textwidth]{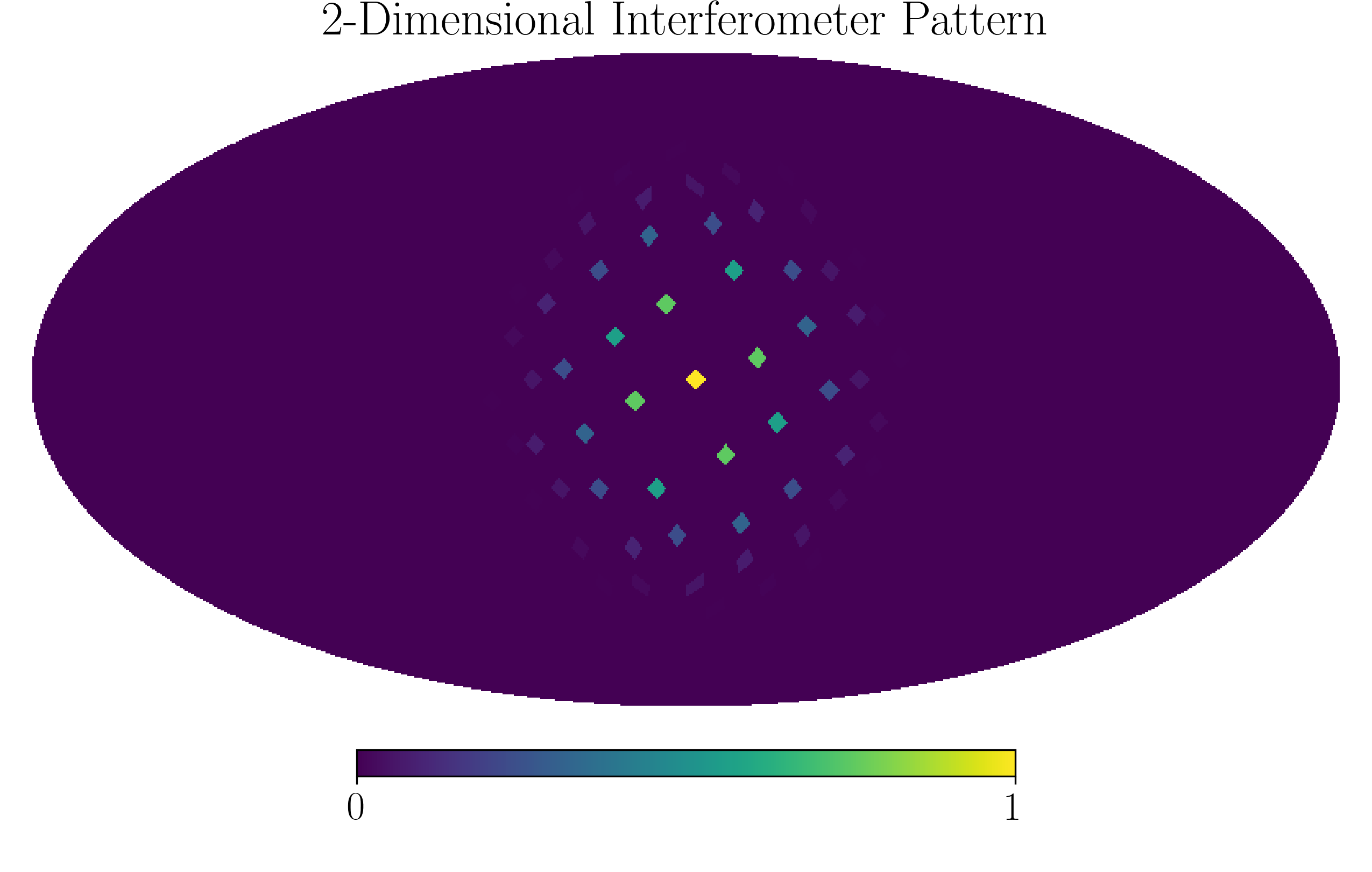}
        \caption{\label{fig:antenna_patterns}Antenna patterns used for simulated map reconstructions. The three panels show the great-circle, 1-D interferometer, and 2-D interferometer patterns described in Section \ref{sec:antenna}. All are shown in HEALPix pixelization with $N_\mathrm{side}=16$ (3072 pixels), although in the calculations below higher resolution was used (see Table \ref{parametertable}).}
    \end{figure*}
    
    \subsection{\label{sec:filter}Filtering the Pure Signal}
    While computing $\mathbf{\bar{M}}^{-1}$ in spherical harmonic space allows us to avoid computing $\mathbf{M}^{-1}$ when finding linear combinations, the computation of $\hat{s}$ in equation (\ref{eq:s}) still relies on $\mathbf{M}$. Because $\mathbf{M}^{-1}$ is the product of sparse matrices, we can use conjugate gradient methods to solve 
        \begin{equation}
        \mathbf{M}^{-1}\hat{s} = \mathbf{A}^T \mathbf{N}^{-1} \vec{d}
    \end{equation}
    for $\hat{s}$
    and hence for the optimal wavelength combinations $\hat{u}$ efficiently.
    
Unfortunately, $\mathbf{M}^{-1}$ has a null space of modes to which the observations are completely insensitive. For example, consider the monopole subspace of signals that are independent of position but may depend on wavelength. The pointing matrix treats all wavelengths identically in this subspace -- the wavelength dependence of the antenna pattern is irrelevant for monopole signals. There are thus $N_f-1$ wavelength-dependent monopole maps that lie in the null space of $\mathbf{A}$ and hence of $\mathbf{M}^{-1}$. In fact, for any $l$, only $2l+1$ wavelength combinations can be reconstructed, so for every $l$ such that $2l+1<N_f$, there are $N_f-(2l+1)$ unmeasureable modes. 

Because of the $\mathbf{A}^T$ on the right side of equation (\ref{eq:s}), the right side lies in this null space and the conjugate gradient solution proceeds without issue. In interpreting the resulting signal estimator, one must remember that these modes have been filtered out of it. In particular, when we compare reconstructed maps with a hypothetical input signal, we must filter the input maps. To be specific, we can compute a filtered input signal map $\bar{s}$ from a candidate signal map $\vec s$
by finding a conjugate gradient solution to
    \begin{equation}
        (\mathbf{A}^T \mathbf{N}^{-1} \mathbf{A})\bar{s} = \mathbf{A}^T \mathbf{N}^{-1} (\mathbf{A} \vec{s}).
        \label{eq:sbar}
    \end{equation}
This process removes the undetectable modes from the pure signal, allowing for a fair comparison of $\hat{s}$ and $\bar{s}$. A visual representation of the pure, filtered, and estimated signals can be found in Section \ref{sec:results}.
    
    \section{\label{sec:results}Results} 
    
    \subsection{Antenna patterns}\label{sec:antenna}
    We tested three different asymmetric antenna patterns to produce simulated data in order to perform this analysis for polychromatic map-making. These are, in order of increasing complexity, a great-circle antenna pattern, 1-dimensional interferometer antenna pattern, and a 2-dimensional interferometer antenna pattern, as shown in Fig. \ref{fig:antenna_patterns}. Here and throughout this paper, all pixelized data on the sphere are  in HEALPix pixelization \cite{healpix}, and all computations on such maps use the HEALPy Python implementation of HEALPix \cite{healpy}.
    
    We emphasize that our goal is not to simulate actual instruments in full detail (although the 2-D interferometer may be considered ``QUBIC-like'' \cite{qubic1,qubic2}, and the great-circle pattern is ``CHIME-like" \cite{chime}). Rather, our goal is to show illustrative examples of asymmetric antenna patterns of the sort to which the methods described in this paper are applicable.
    
    The great-circle pattern is that produced by a telescope that has much sharper resolution along one direction than along the orthogonal direction (as might be produced by a parabolic dish that is long but not wide). In the flat-sky approximation, its antenna pattern would be of the form $A(x,y) = \exp[-x^2/(2\sigma_x^2)-y^2/(2\sigma_y^2)]$ with $\sigma_x\gg\sigma_y$. As noted in Section \ref{sec:mono}, no useful information can be recovered on scales smaller than $\sigma_y$, so we take the signal vector we are attempting to reconstruct to have been smoothed by a Gaussian with beam width $\sigma_y$. In this case, the antenna pattern has zero width in the $y$ direction and width $\sigma=\sqrt{\sigma_x^2-\sigma_y^2}$ in the $x$ direction. The antenna pattern is thus a Gaussian of width $\sigma$ lying along a great circle, with delta-function behavior in the perpendicular direction.
    
    In the one- and two-dimensional interferometer patterns, we imagine an antenna pattern with a series of equally-spaced peaks lying under a broad Gaussian envelope. The peaks are all assumed to have identical shapes, and once again we take the reconstructed signal to have been smooothed by these shapes, allowing us to replace each peak with a delta function. The two-dimensional interferometer pattern is thus similar to what would be found in an interferometer with a square array of antennas.
    
    In all cases, the antenna patterns scale linearly with wavelength -- that is, the widths of Gaussians and the spacing between peaks are all proportional to $\lambda_f$. Since this scaling is the only way wavelength appears in any of our computations, we express $\lambda_f$ in units that make the descriptions of our antenna patterns as simple as possible.
    
    For the Great Circle pattern, we assume a Gaussian beam and measure wavelength in units such that  the beam width along the great circle is $\lambda_f$. 
    For the interferometer patterns, we measure wavelength in units such that the space between peaks is exactly $\lambda_f$. The heights of the peaks are determined by the beam pattern of a single antenna, which is of course broader than the individual peaks. We take this pattern to be Gaussian with width $1.32\lambda_f$, meaning that the first peak away from the center has height 0.5 times the central peak. When modeling the antenna patterns, we keep all peaks with amplitude at lest $10^{-3}$ times the central peak.

\begin{table*}
        \begin{tabular}{|c|c|c|c|c|c|c|}
            \hline
            \multicolumn{3}{|c|}{\ }&\multicolumn{2}{|c|}{Full Sky}& 
            \multicolumn{2}{|c|}{Half Sky}\\
            \hline
            \hline
            Antenna Pattern & $N_p$ & $N_f$ & $N_t$ & $\sigma_{\vec{s}} / \sigma_t$ & $N_t$ &$\sigma_{\vec{s}}/\sigma_t$  \\ [0.5ex]
            \hline
            2-Dimensional Interferometer & 12288 & 20 & 4\,91520 & 2.717&4\,91520 &2.762  \\
            \hline
            1-Dimensional Interferometer & 12288 & 14 & 3\,44064 & 3.021&17\,20320 & 3.077 \\
            \hline
            Great Circle & 12288 & 18 & 4\,42368 & 2.463&4\,42368 & 3.400 \\
            \hline
        \end{tabular}
        \caption{Map parameters. For each of the three antenna patterns described in Section \ref{sec:results}, we give the number of pixels $N_p$ and the number of frequency bins $N_f$ in the simulations. For both the full-sky and half-sky simulations, we give the number of time steps $N_t$ and the signal-to-noise ratio. In the latter, $\sigma_{\vec{s}}$ is the standard deviation of the signal vector, and $\sigma_t$ is the standard deviation of the  noise in the time domain. The choice of values for the various parameters is discussed in Sections \ref{sec:simsignals} and \ref{sec:harmonic}.}
         \label{parametertable}
    \end{table*}

\subsection{Simulated signals}\label{sec:simsignals}
    Throughout this section, the simulated signal vector, $\vec{s}$, is composed of linear combinations of two independent maps, with wavelength-dependent weights. To be specific, let $f$ and $g$ be two Gaussian random fields on the sphere. Then the signal at any location $\hat{r}$ and wavelength $\lambda$ is a linear interpolation of these two functions:
    \begin{equation}
        s(\hat{r};\lambda) = wf(\hat{r})+(1-w)g(\hat{r}),
    \end{equation}
    where $w = (\lambda-\lambda_\mathrm{min})/(\lambda_\mathrm{max}-\lambda_\mathrm{min})$ ranges from 0 at the minimum wavelength in the observed band to 1 at the maximum wavelength.
    
    The pointing matrix is applied to this signal, and then random noise is added to that product to create $\vec{d}$. The noise vector, $\vec{n}$, is created by generating $N_t$ independent Gaussian random numbers with mean 0 and a user-supplied standard deviation $\sigma_t$. The reconstructed signal $\hat{s}$ is then computed using the conjugate gradient method with the expression described in equation (\ref{eq:s}). 
    
   The parameters used in the maps in this section can be found in Table \ref{parametertable}. In all cases, we use HEALPix resolution $N_{\mathrm{side}}=32$, leading to $N_p = 12288$. We follow the procedure described in the next section to determine the number $N_f$ of frequency sub-bands required for each antenna pattern. Finally, we choose the number of time steps to be $N_t=2N_fN_p$, which insures that the linear system to be solved in the full signal reconstruction is well-conditioned. (Although the full signal reconstruction is not particularly useful due to its large errors, it is still convenient for it to be possible to compute it numerically.)
    
\subsection{Full-sky harmonic space calculations.}\label{sec:harmonic}
    \begin{figure*} 
        \centering
        \subfloat[2-Dimensional Interferometer]{\includegraphics[width=0.33\textwidth]{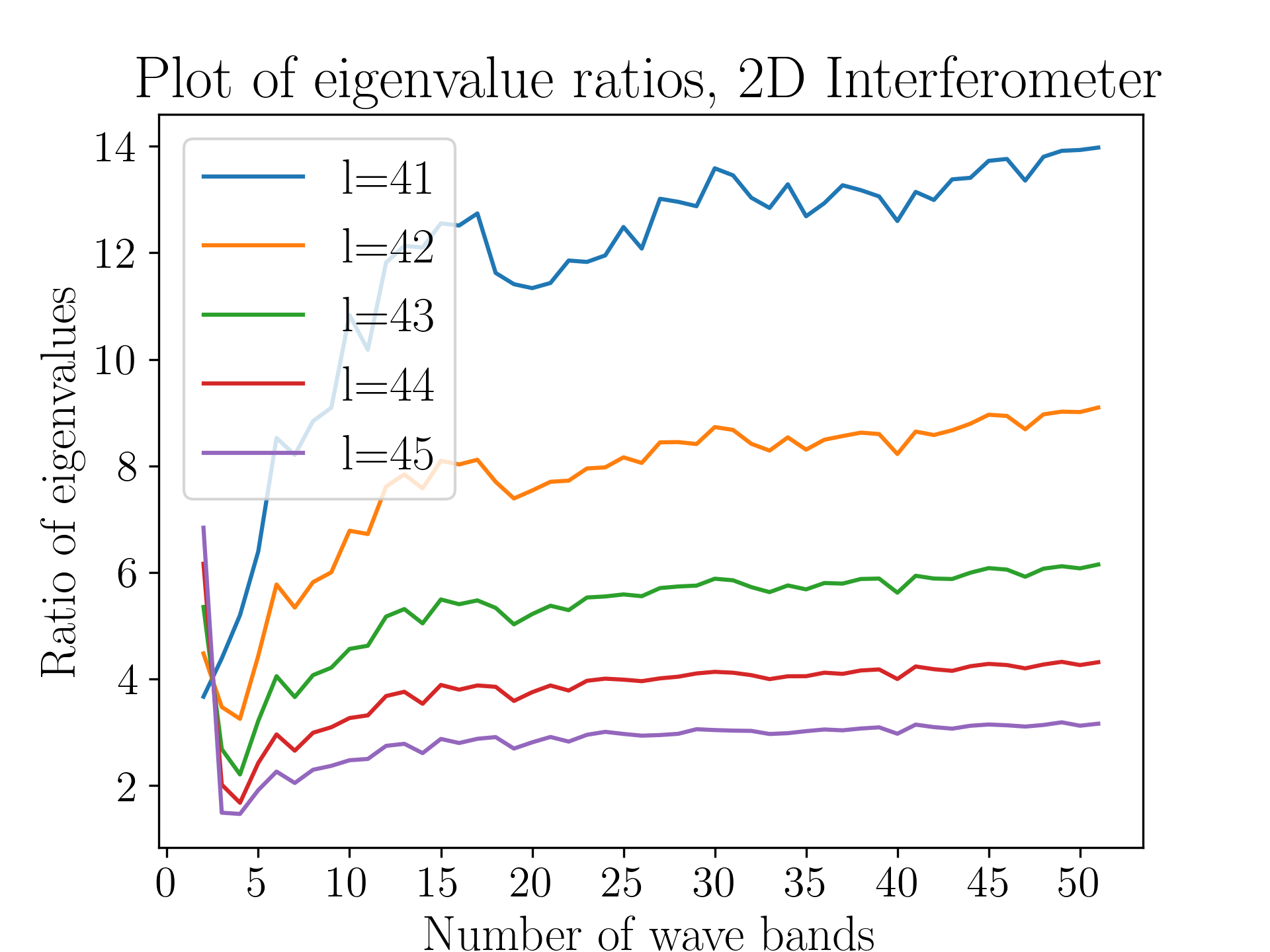}\label{fig:eigenvalues_2D}}
        \subfloat[1-Dimensional Interferometer]{\includegraphics[width=0.33\textwidth]{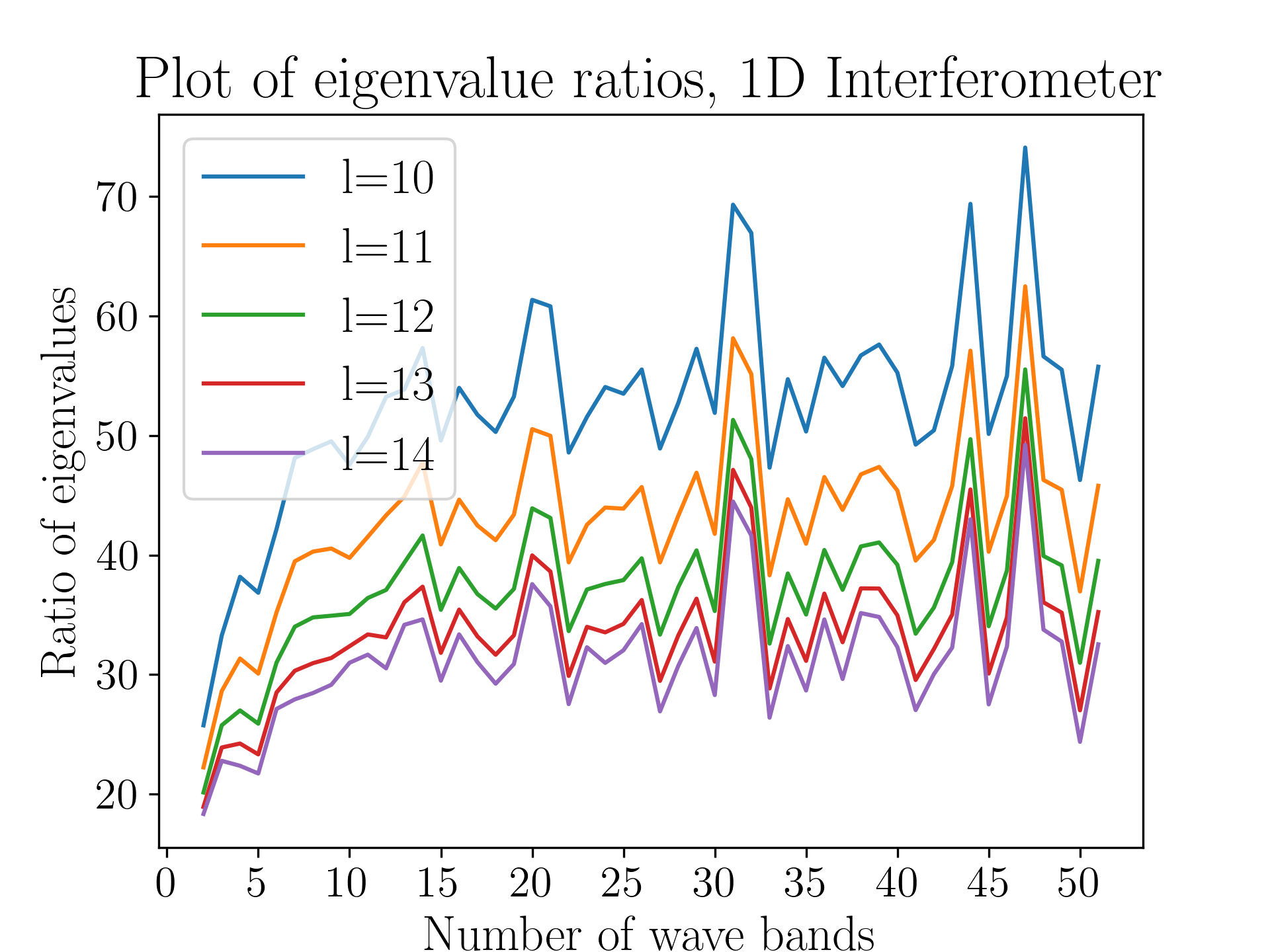}\label{fig:eigenvalues_1D}}
        \subfloat[Great Circle]{\includegraphics[width=0.33\textwidth]{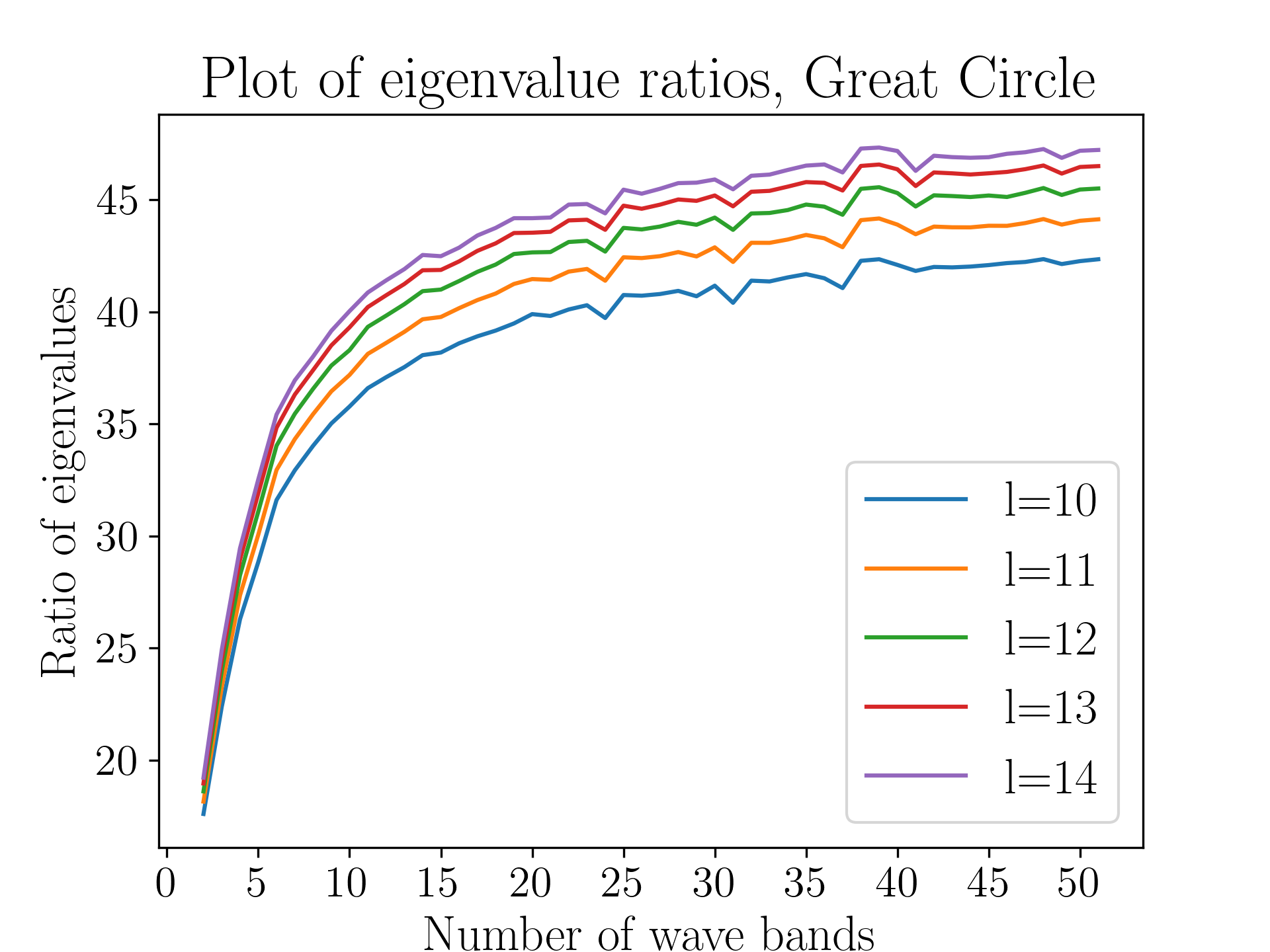}\label{fig:eigenvalues_GC}}
        \caption{Ratio of greatest two eigenvalues of $\mathbf{C}_l$. The number of wavelength bins in the input signal vectors were chosen such that the number of bins would be an approximation of the continuous case. We choose values of $N_f$, listed in Table \ref{parametertable}, to be  large enough that the ratio of the eigenvalues is close to its asymptotic limit. 
        }
         \label{fig:eigenvalues}
    \end{figure*}
   
   We first performed calculations in harmonic space for a hypothetical full-sky experiment with isotropic random sampling of antenna orientations. 
   
   We begin by considering the choice of $N_f$, the number of wavelength bins in our input signal vector. We wish effects of wavelength discretization to be unimportant, so we choose $N_f$ to be large enough to produce a reasonable approximation of the continuous case. In particular, as $N_f$ increases, the eigenvalues and eigenvectors of $\mathbf{C}_l$ should approach stable continuum limits. We use the ratio of the two highest eigenvalues as a measure of this approach to stability. These ratios are plotted in Fig. \ref{fig:eigenvalues}. Based on these plots, we chose the values of $N_f$ in Table \ref{parametertable}.
    
    \begin{figure*}
        \centering
        \subfloat[2-Dimensional Interferometer]{\includegraphics[width=0.33\textwidth]{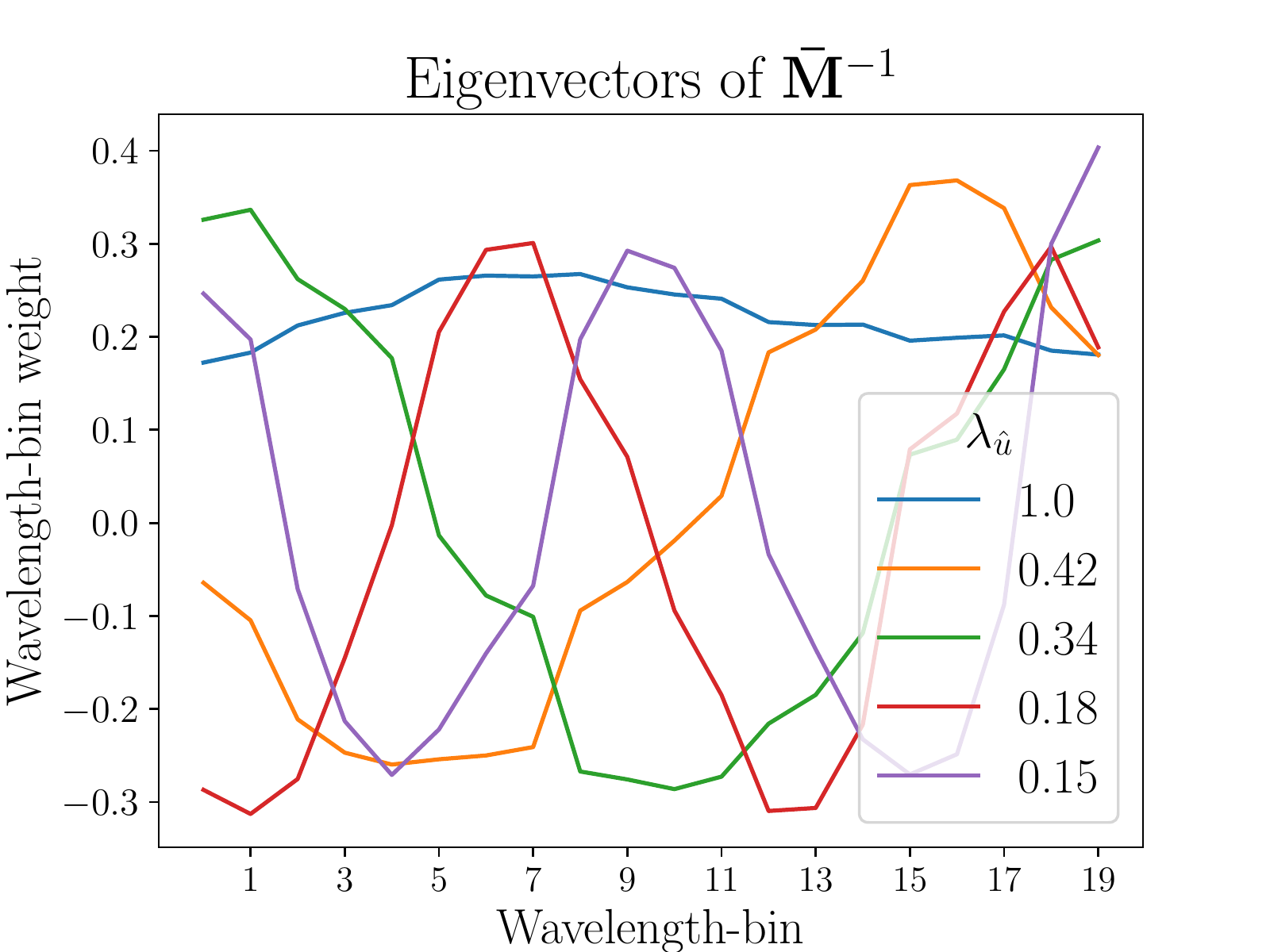}\label{fig:eigenvectors_2D}}
        \subfloat[1-Dimensional Interferometer]{\includegraphics[width=0.33\textwidth]{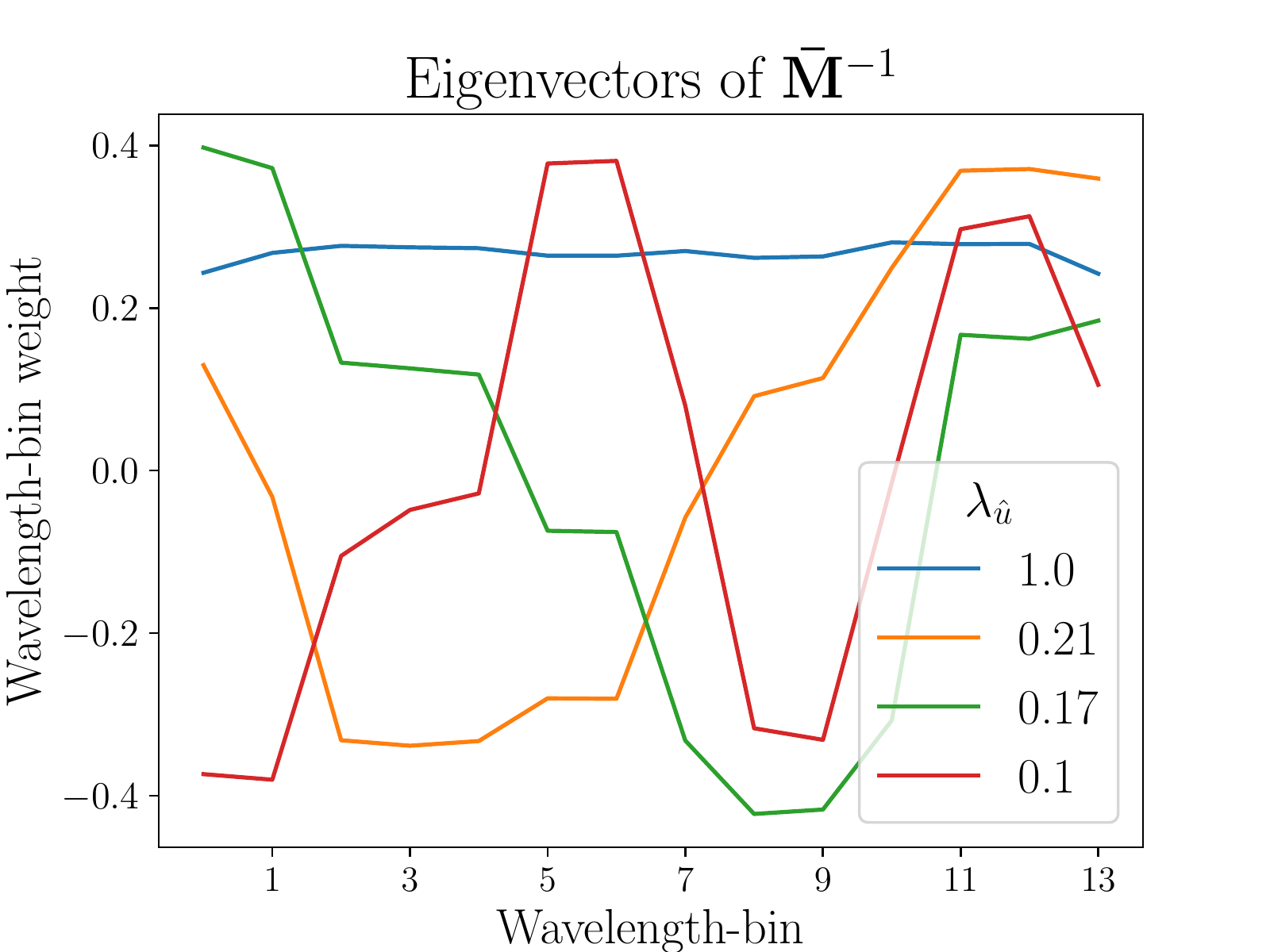}\label{fig:eigenvectors_1D}}
        \subfloat[Great Circle]{\includegraphics[width=0.33\textwidth]{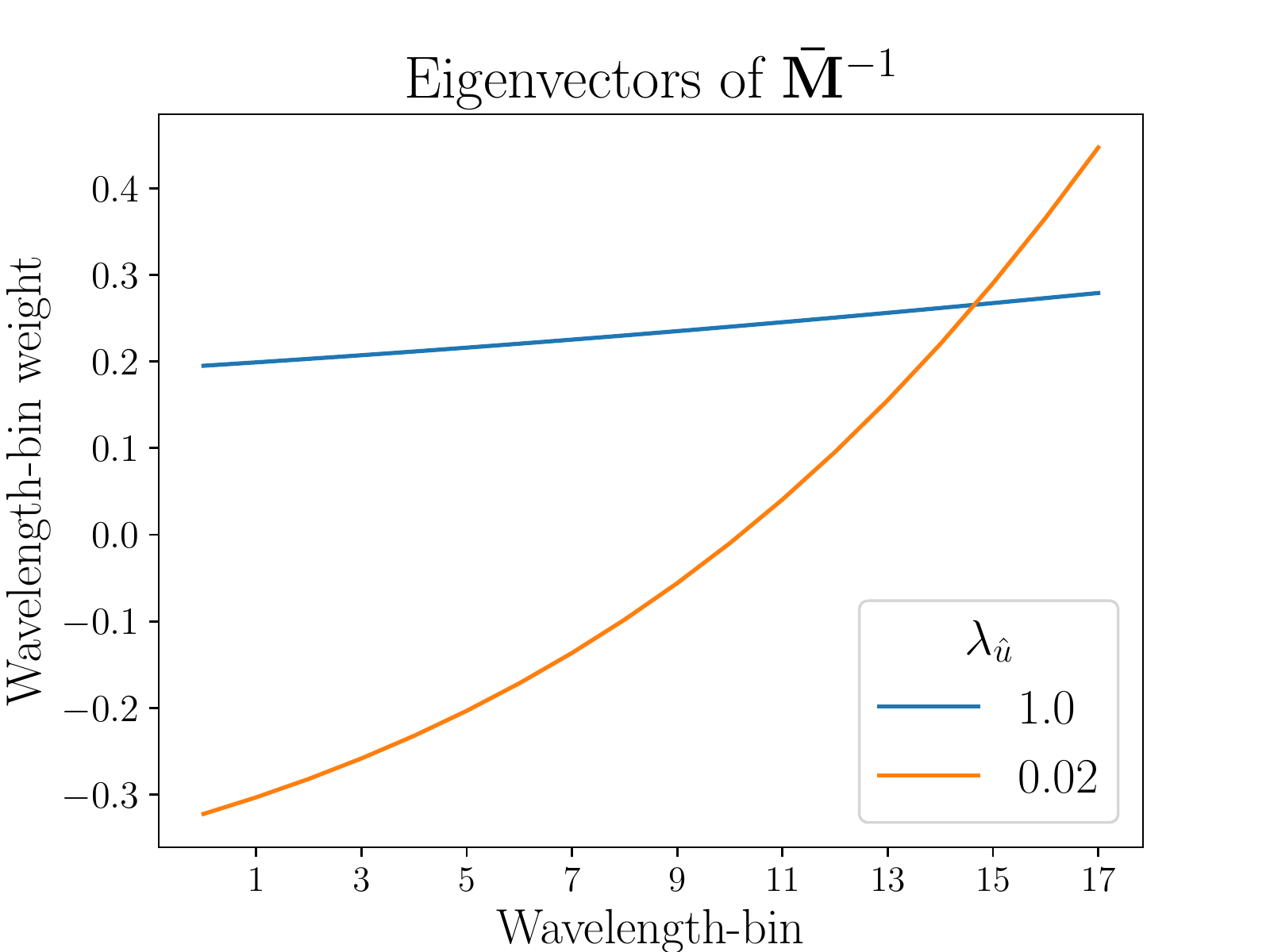}\label{fig:eigenvectors_GC}}
        \caption{Eigenvectors of $\mathbf{\bar{M}}^{-1}$ for each antenna pattern, assuming an isotropic experiment.} \change{As described in Section \ref{method}, the eigenvectors are the optimal colors to reconstruct.} The horizontal axis is the bin number in wavelength, with 0 corresponding to $\lambda_\mathrm{min}$ and the maximum value to $\lambda_\mathrm{max}$. The legend gives the eigenvalue, scaled by the maximum eigenvalue.
         \label{fig:eigenvectors}
    \end{figure*}
    
    We can immediately note from Fig. \ref{fig:eigenvalues} that the two-dimensional interferometer is far more promising than the other two antenna patterns for polychromatic map-making: the smaller eigenvalue ratios mean that the uncertainties in the second-best map will be closer to those in the best map in this case than in the other two.
    
    Fig. \ref{fig:eigenvectors} shows eigenvectors of $\mathbf{\bar{M}}^{-1}$ for the various antenna patterns. For the two interferometer patterns, the eigenvectors plotted have eigenvalues that are at least 0.1 times the largest eigenvalue, meaning that the noise levels in these maps will be worse than those in the best map by at most a factor of $\sqrt{10}$. For the great-circle case, the second-best eigenvector is shown even though it lies below this threshold.
    
    It may come as no surprise that the linear combination corresponding to the smallest uncertainty is reminiscent of a total-intensity map, with weights approximately equal at all wave-bands. Thus, the bulk of the additional information available via polychromatic map-making (as compared to monochromatic) is found in the second-largest eigenvector, which represents a color map.

    \begin{figure*}
        \centering
        \subfloat[Pure signal, $\vec{s}$]{\includegraphics[width=0.33\textwidth]{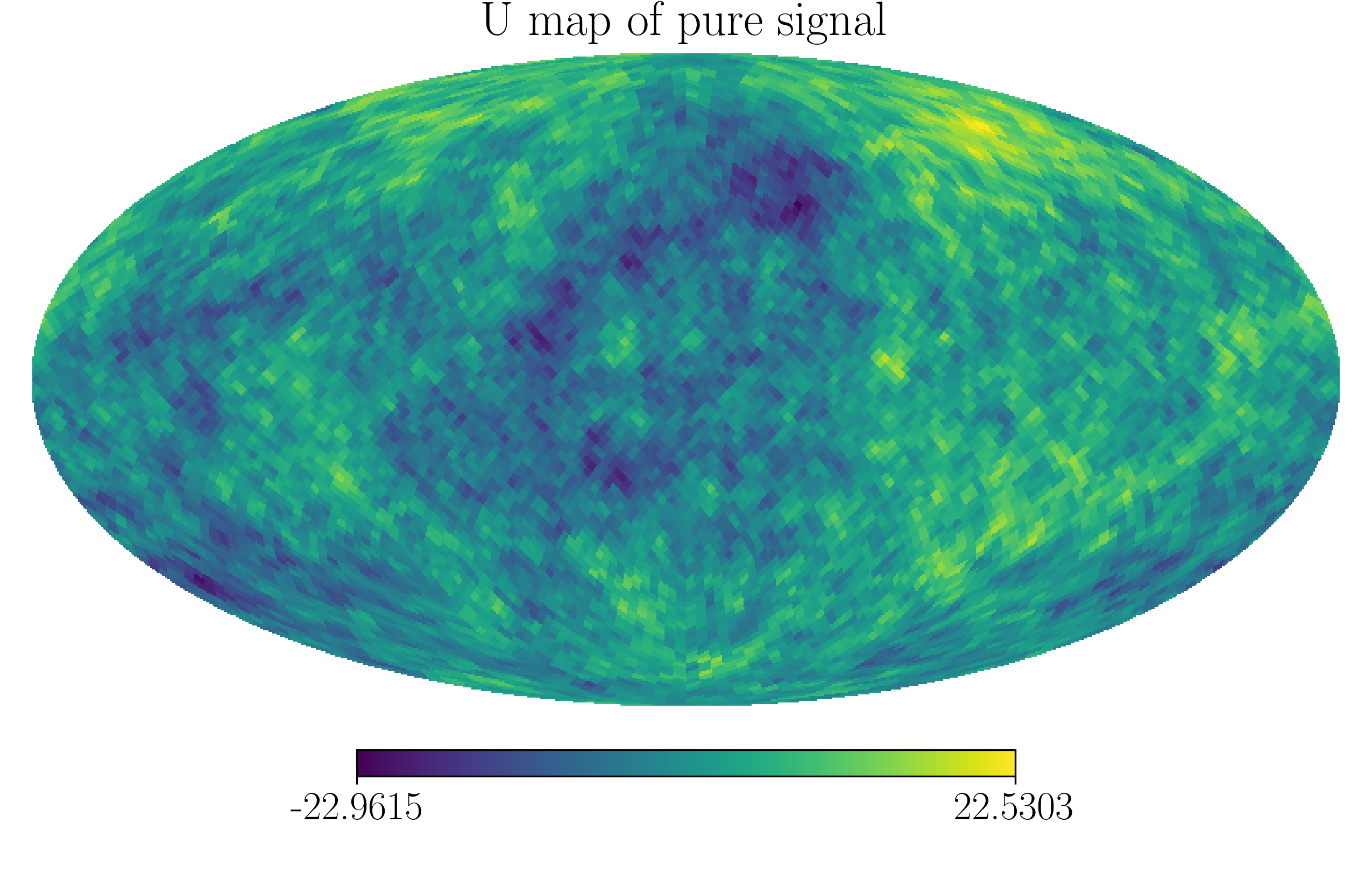}\label{fig:upure_2D}}
        \subfloat[Estimated Signal, $\hat{s}$]{\includegraphics[width=0.33\textwidth]{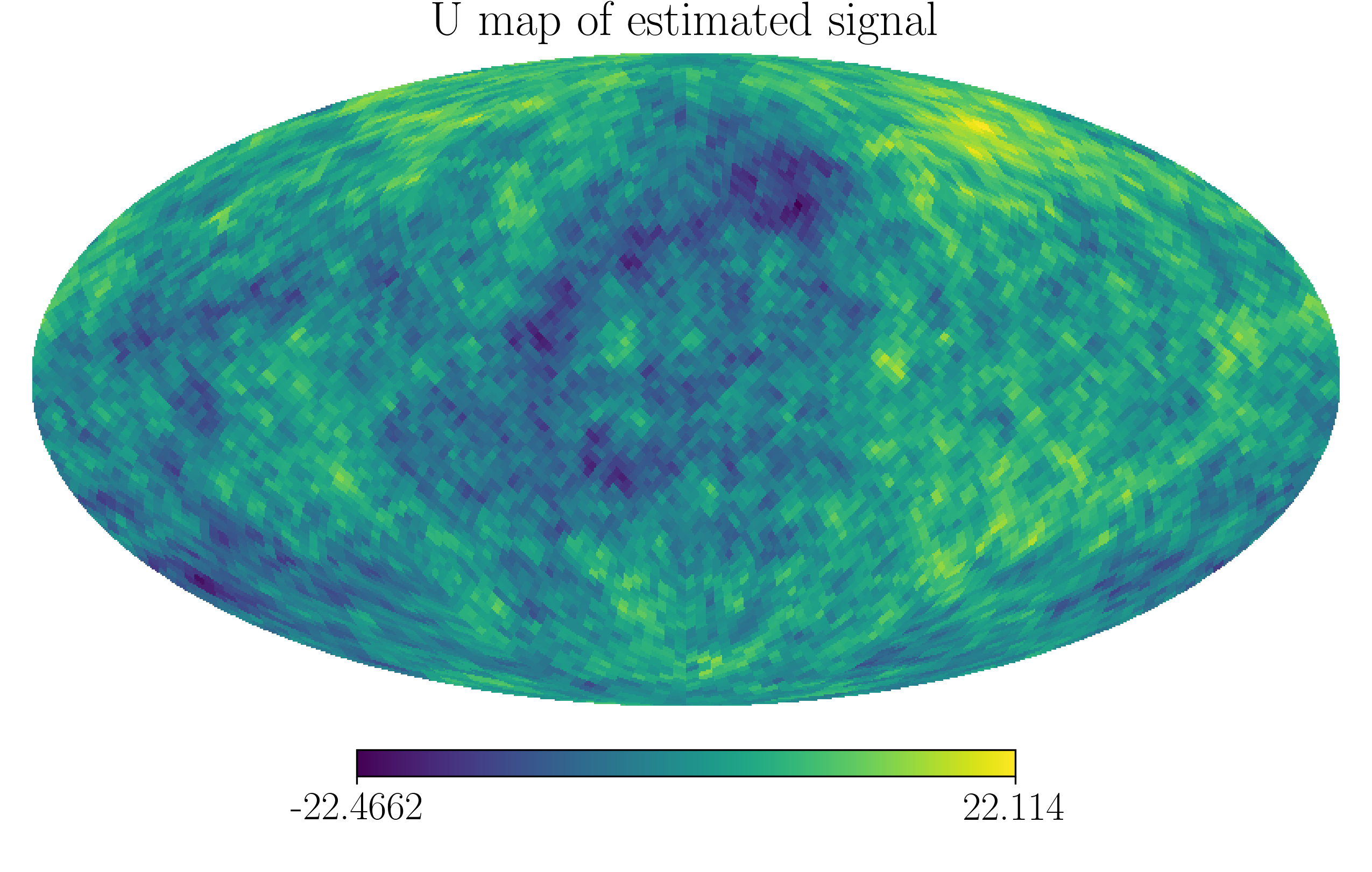}\label{fig:uestimated_2D}}
        \caption{Best linear combination of wavelengths from a simulation using the 2-Dimensional Interferometer pattern. The notation \textit{U map} refers to a signal map in a selected linear combination, such as in the notation of equation (\ref{eq:u_map}). The best linear combination corresponds to a near-total-intensity map, which is essentially the same as what is obtained by monochromatic mapmaking.}
    \end{figure*}
    \begin{figure*}
        \centering
        \subfloat[Pure signal, $\vec{s}$]{\includegraphics[width=0.33\textwidth]{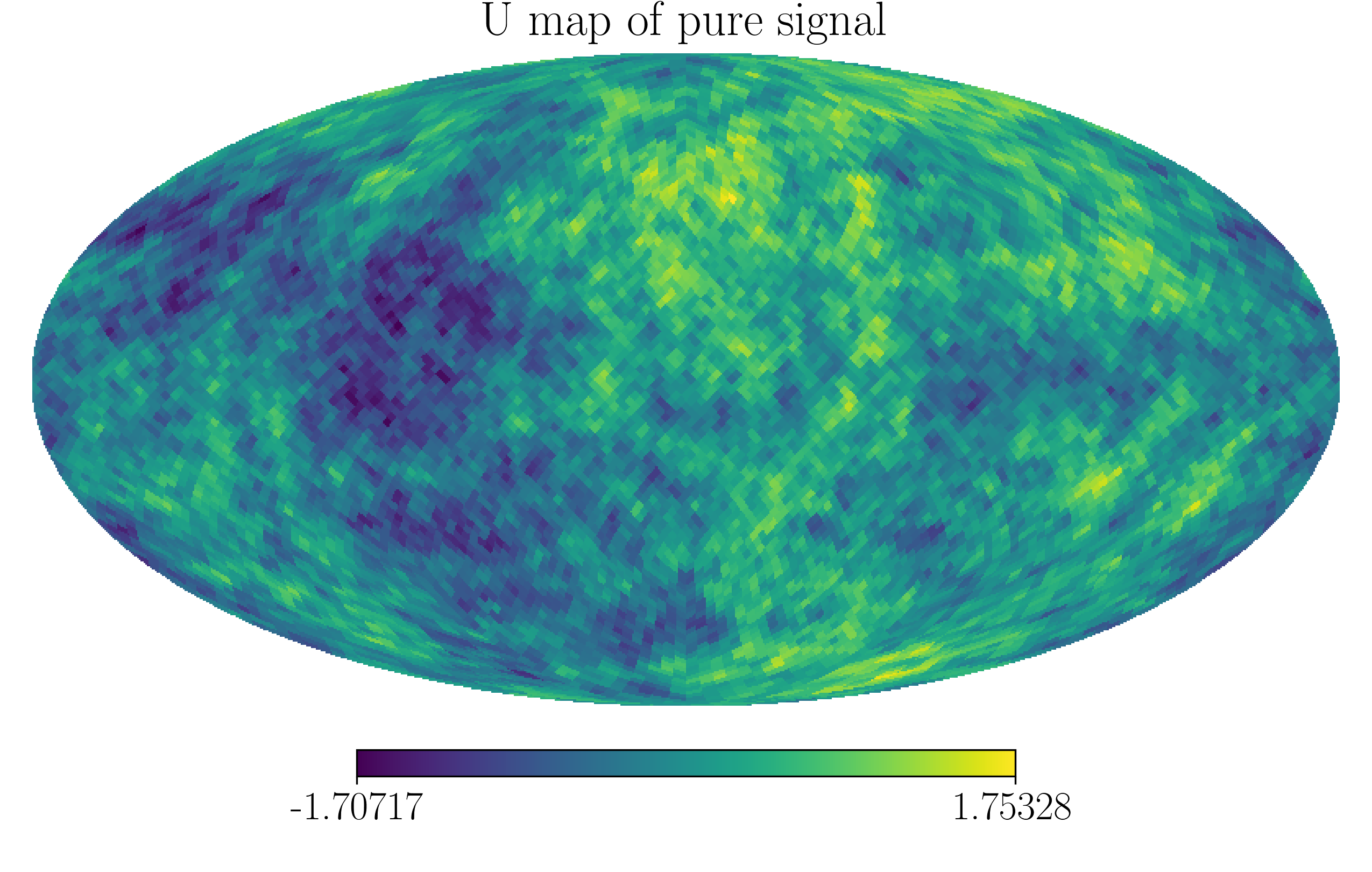}\label{fig:upure_2D1}}
        \subfloat[Estimated Signal, $\hat{s}$]{\includegraphics[width=0.33\textwidth]{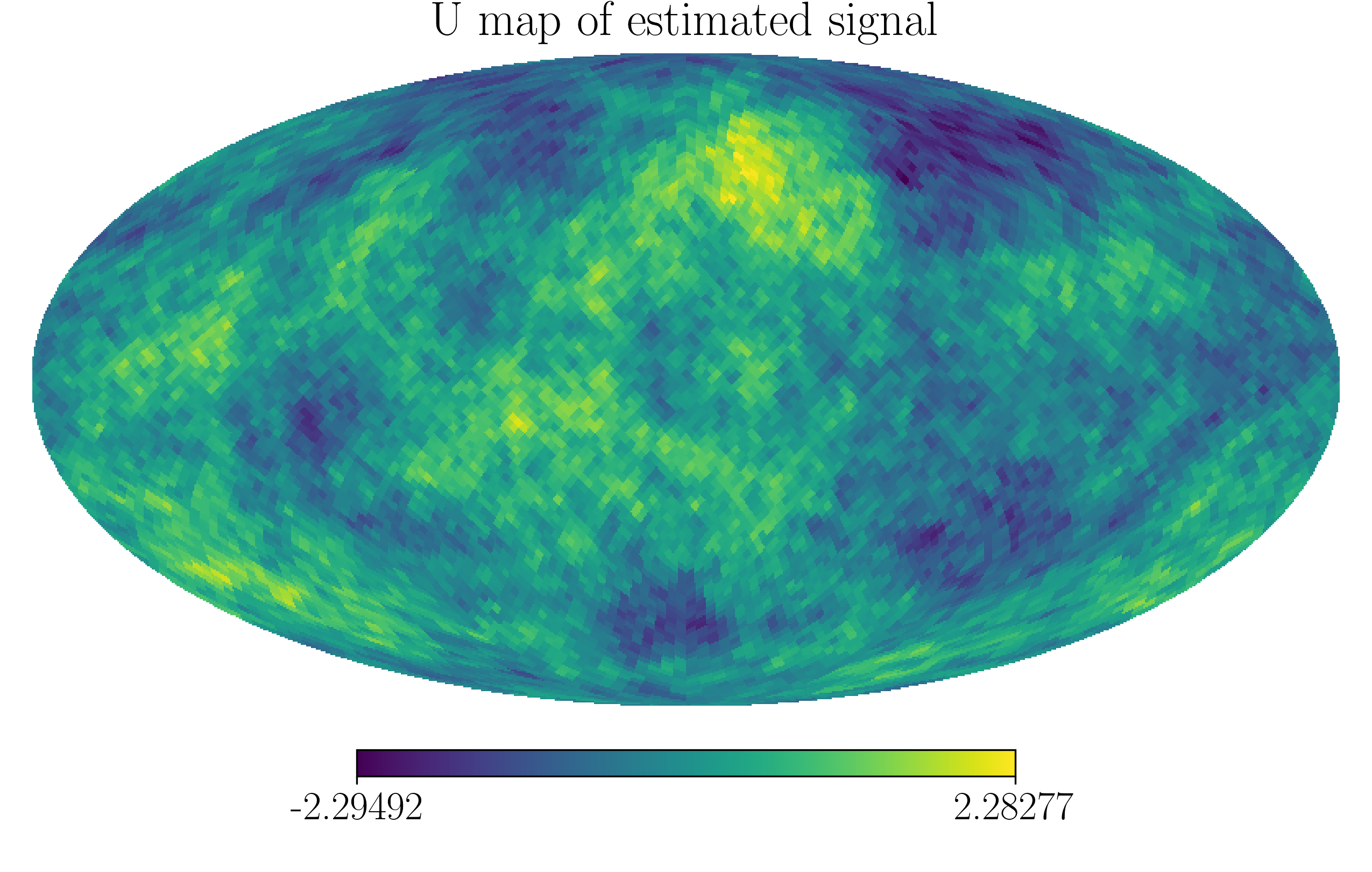}\label{fig:uestimated_2D1}}
       \subfloat[Difference map  $\hat{s}-\vec{s}$]{\includegraphics[width=0.33\textwidth]{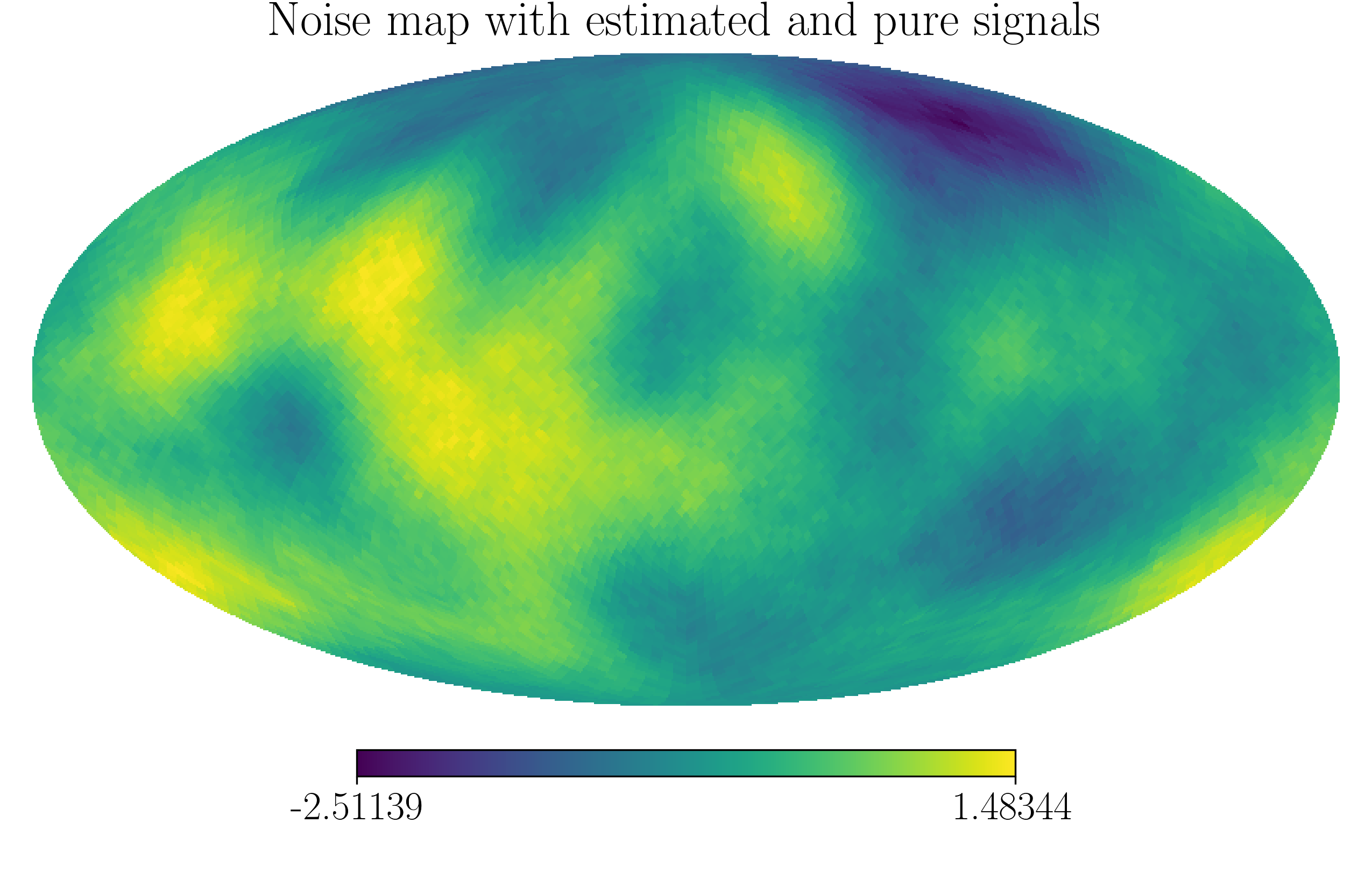}\label{fig:dif_pure}}

       \subfloat[Pure filtered signal, $\bar{s}$]{\includegraphics[width=0.33\textwidth]{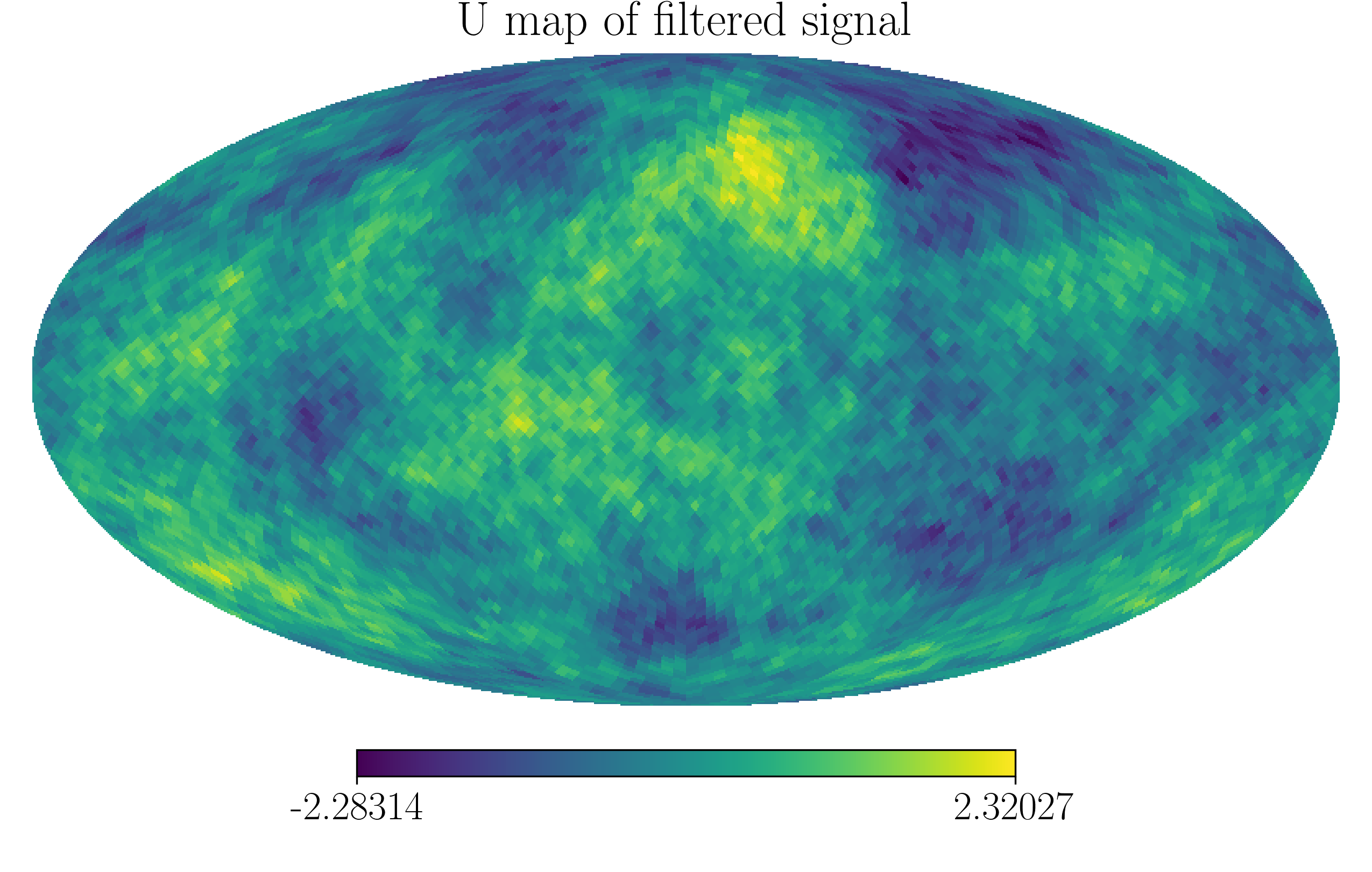}\label{x}}
          \subfloat[Difference map $\hat{s}-\bar{s}$]{\includegraphics[width=0.33\textwidth]{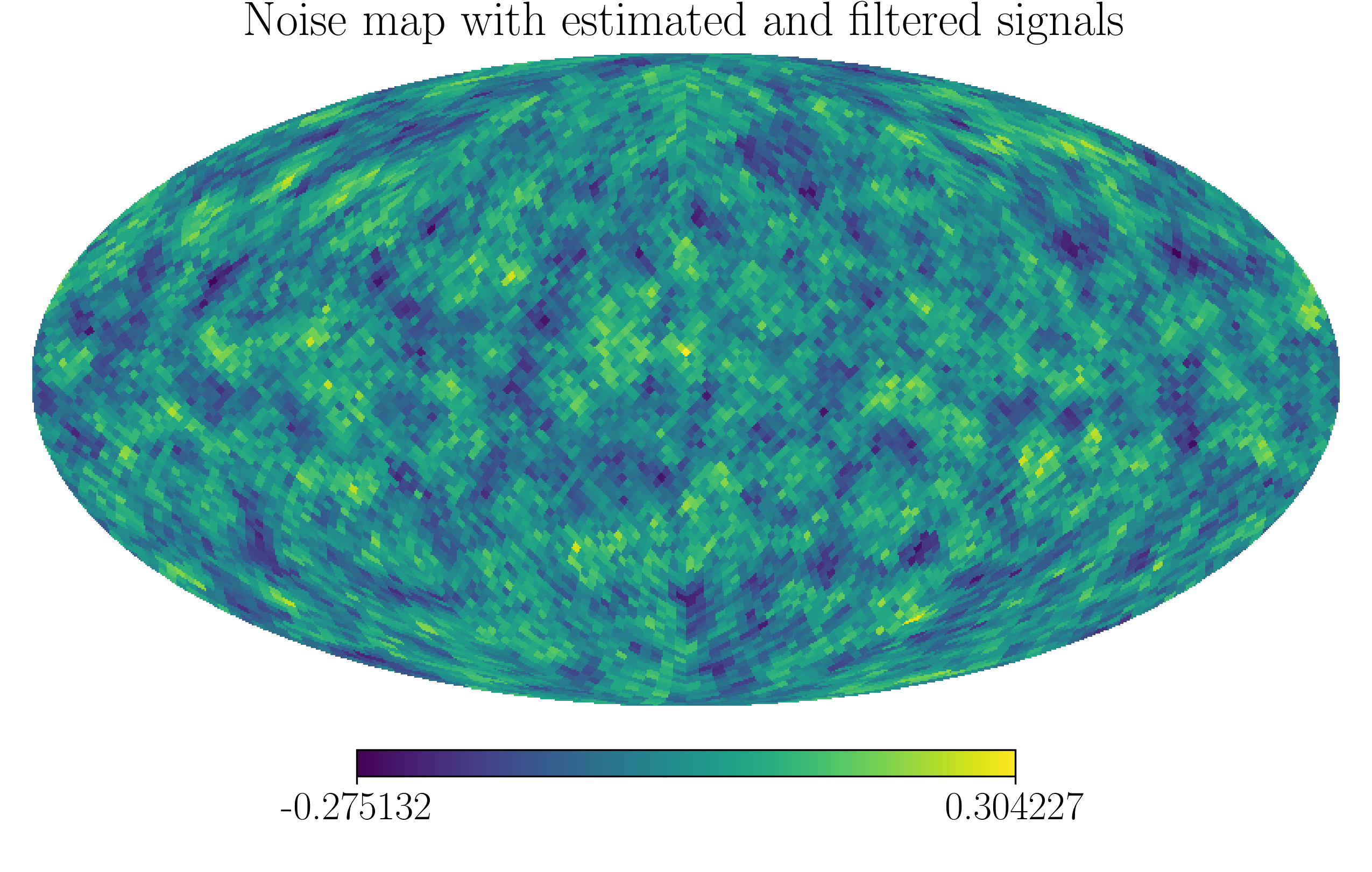}\label{fig:dif_filtered}}
        \caption{\label{fig:secbest_2D}Second-best linear combination of wavelengths from a simulation using the 2-Dimensional Interferometer pattern. \change{The difference map $\hat{s} - \vec{s}$ shows the modes in the null space of $\mathbf{M}^{-1}$ that must be filtered (see Section \ref{sec:filter}.). The difference map $\hat{s} - \bar{s}$ lacks the large-scale power of the original difference map, which is expected. See eq.\ (\ref{eq:sbar}) for the definition of $\overline{s}$.}}
        \label{fig:secondbest2d}
    \end{figure*}
    
    \subsection{Full-sky maps in pixel space}
    
    While the harmonic-space calculations above tell us the expected noise level in the reconstructions, it is  helpful to view actual maps of the signals. Fig. \ref{fig:upure_2D} and \ref{fig:uestimated_2D} show maps of the best linear combination of wavelengths of $\vec{s}$ and $\hat{s}$ for the 2-Dimensional Interferometer pattern. The eigenvector containing the best linear combination from this simulation is the same as the one shown in Fig. \ref{fig:eigenvectors_2D}. The notation \textit{U map} in the figure titles refers to a signal map in a selected linear combination, such as in the notation of equation (\ref{eq:u_map}). 
   
    As mentioned earlier, these near-total-intensity maps are essentially the same as those obtained by monochromatic mapmaking; to evaluate the polychromatic method we must look at the second-best linear combination, that is, to
     the eigenvector with the second-largest eigenvalue in Fig. \ref{fig:eigenvectors_2D}.
     
        Fig. \ref{fig:secbest_2D} shows maps of this combination. It is clear from panels (a) and (b) that the pure signal and estimated signal have discrepancies in structure on a large scale. Panel (c) confirms this impression by showing the difference between the two.
        
        This difference is due to the null space of $\mathbf{M}^{-1}$ described in Section \ref{sec:filter}: modes in the null space that are present in the original signal are absent in the reconstructed map. If we filter the true signal $\vec s$ by removing the null space of $\mathbf{M}^{-1}$ before taking the linear combination, we get the map shown in panel (d). The difference between this and the reconstructed map [panel (e)] is smaller and lacks the large-scale power of the original difference map. 
        
    \begin{figure*} 
        \centering
        \subfloat[Estimated Signal, $\hat{s}$]{\includegraphics[width=0.33\textwidth]{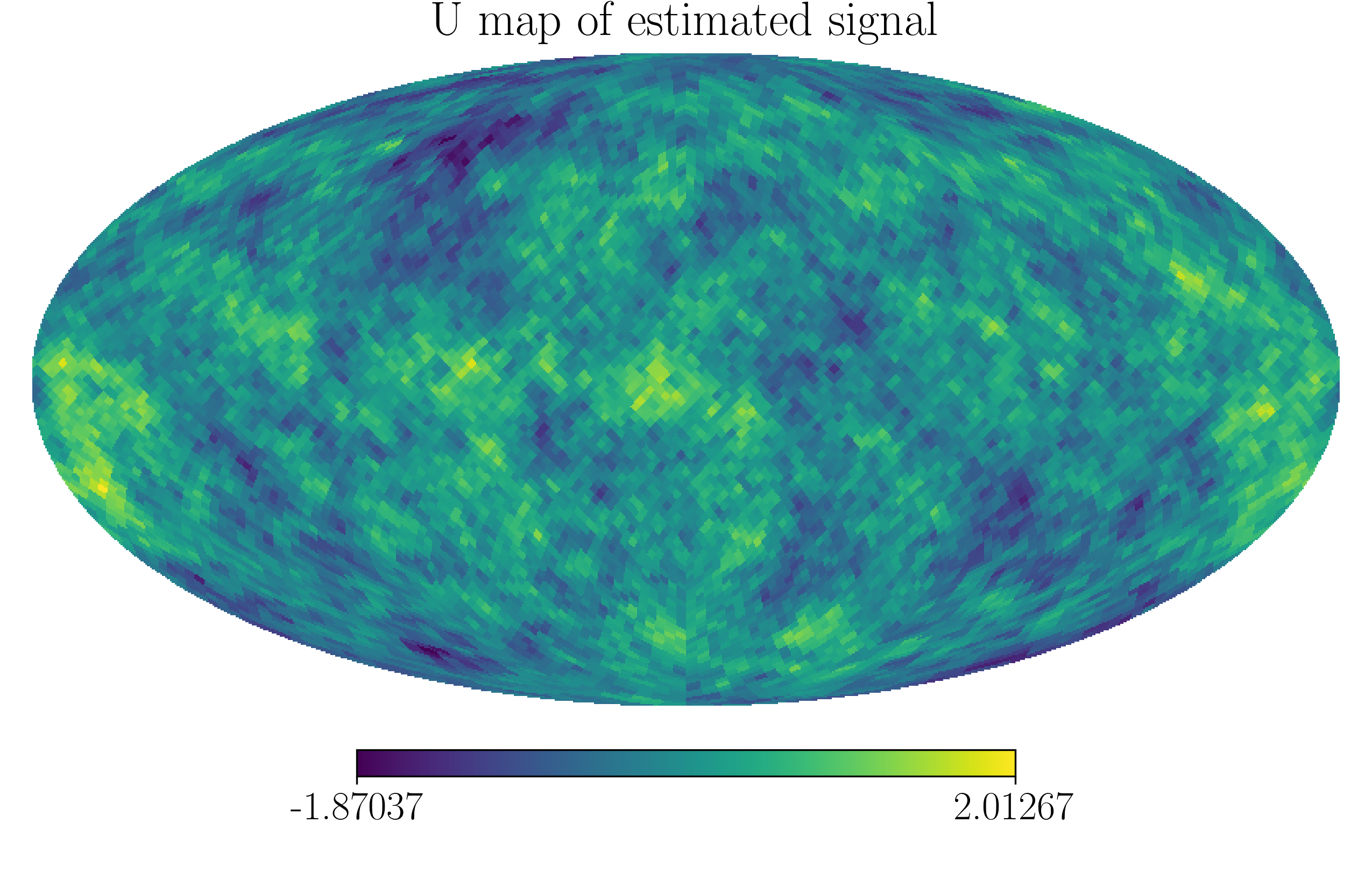}\label{fig:uestimated_1D}}
        \subfloat[Pure filtered signal, $\bar{s}$]{\includegraphics[width=0.33\textwidth]{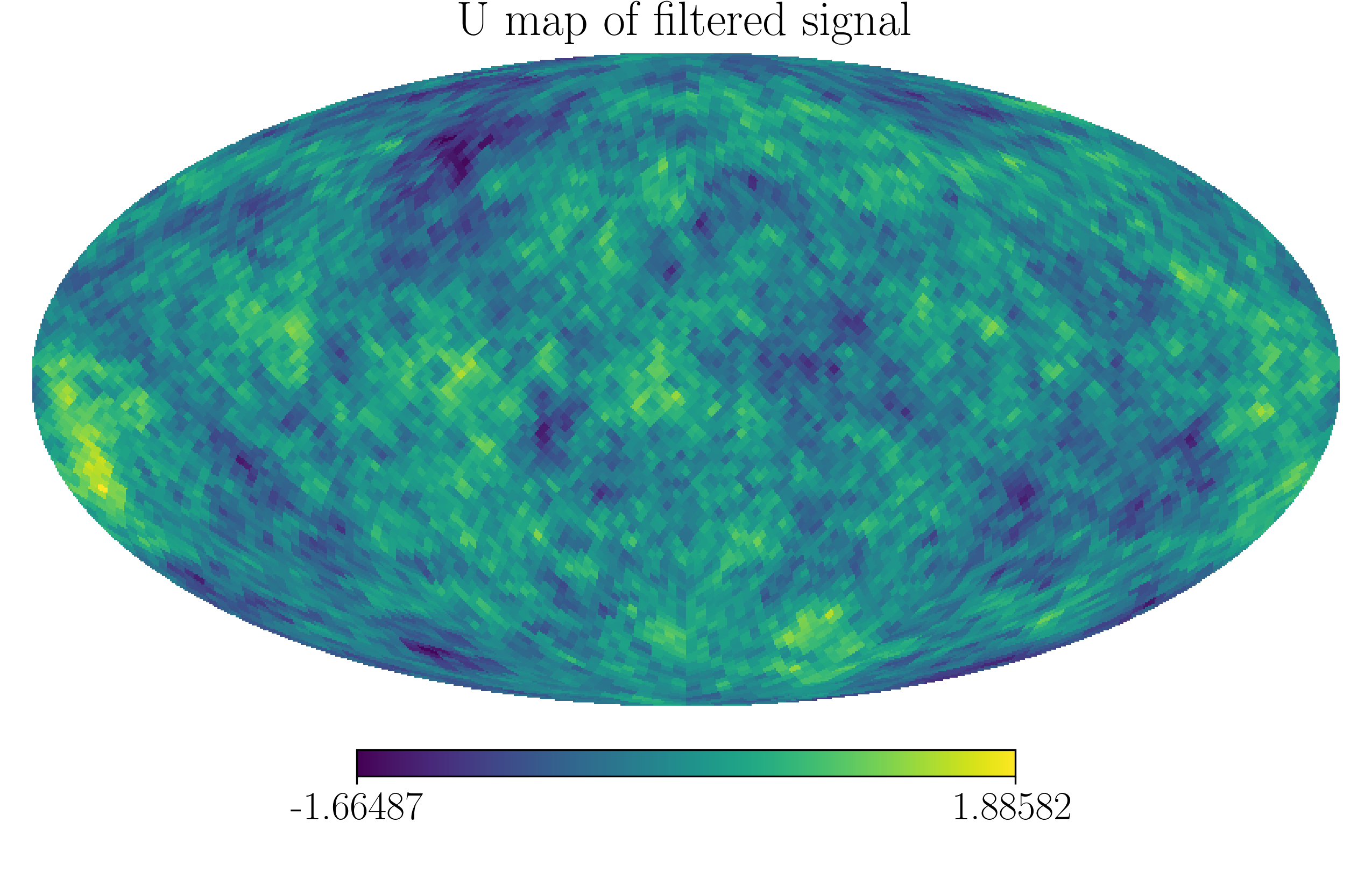}\label{fig:ufiltered_1D}}
        \subfloat[Difference map $\hat{s}-\bar{s}$]{\includegraphics[width=0.33\textwidth]{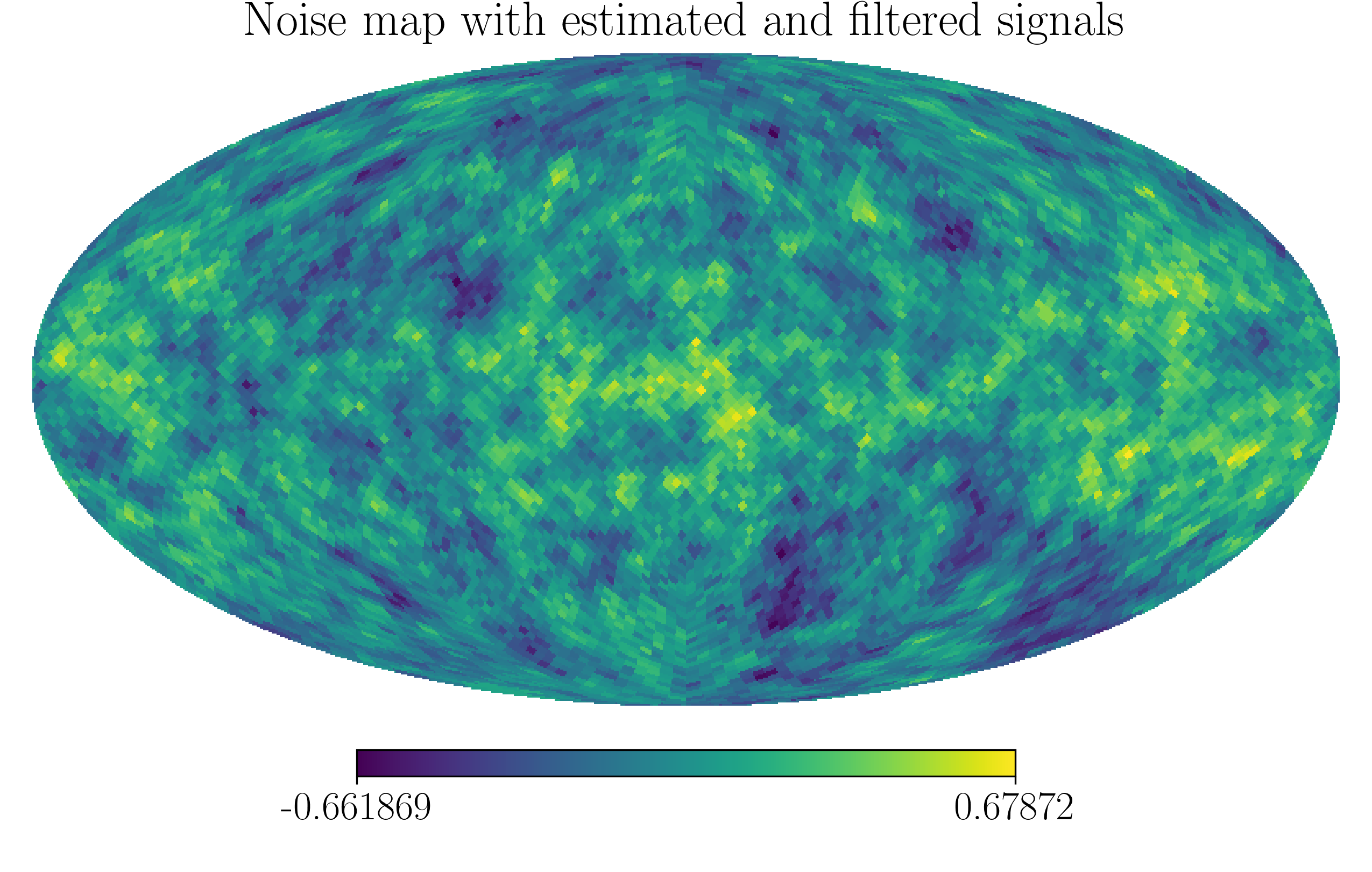}\label{fig:dif_map_1D_full}}
        \caption{\label{fig:1D}Second-best linear combination of wavelengths from a simulation using the 1-Dimensional Interferometer pattern. \change{As in the case of the 2-Dimensional Interferometer shown in Fig.\ \ref{fig:secondbest2d}, there are large differenes in the  unfiltered maps. We omit these plots in this figure, showing only the filtered comparison.}}
    \end{figure*}
   
    \begin{figure*}
        \centering
        \subfloat[Estimated Signal, $\hat{s}$]{\includegraphics[width=0.33\textwidth]{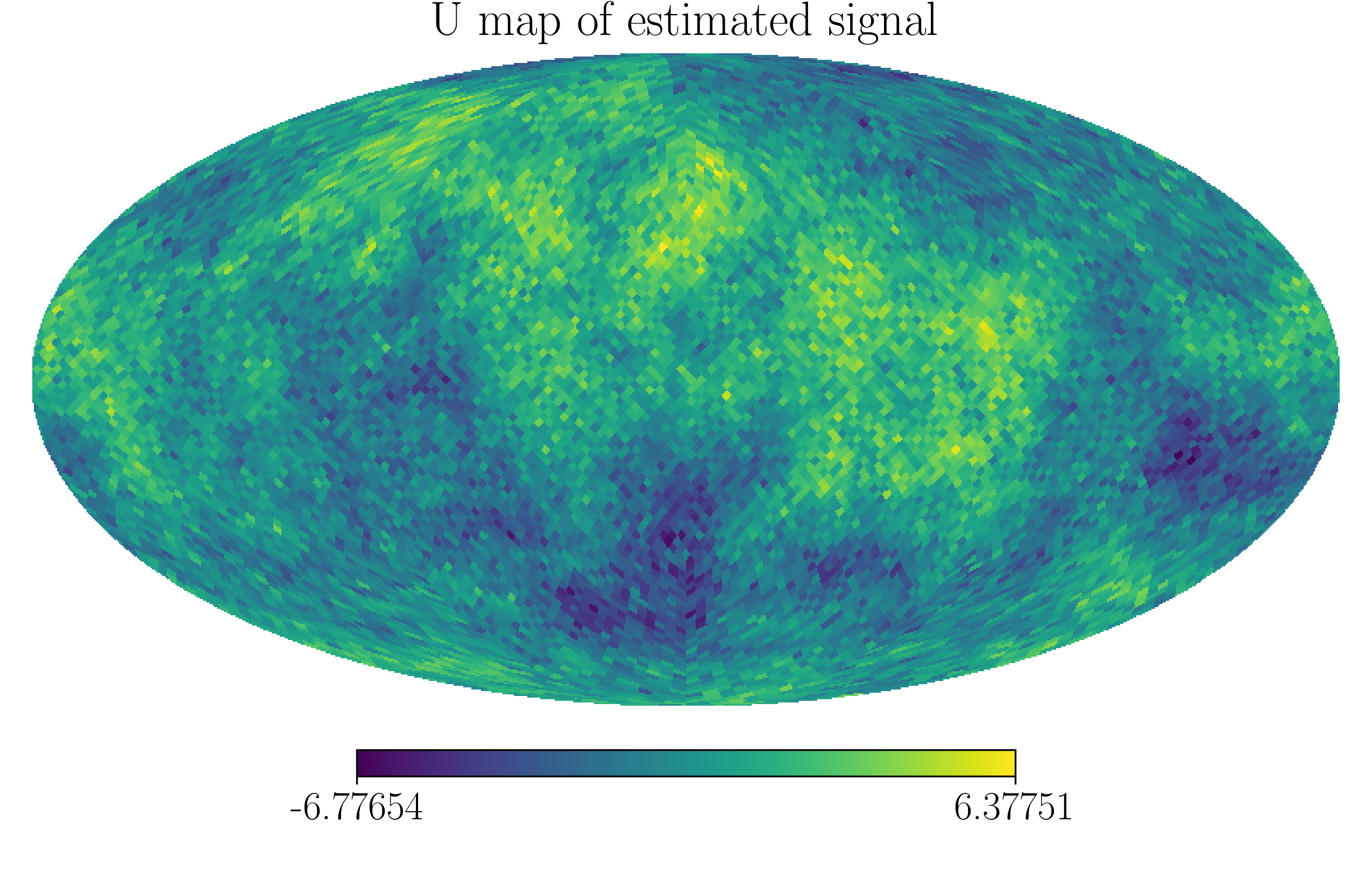}\label{fig:uestimated_GC}}
        \subfloat[Pure filtered signal, $\bar{s}$]{\includegraphics[width=0.33\textwidth]{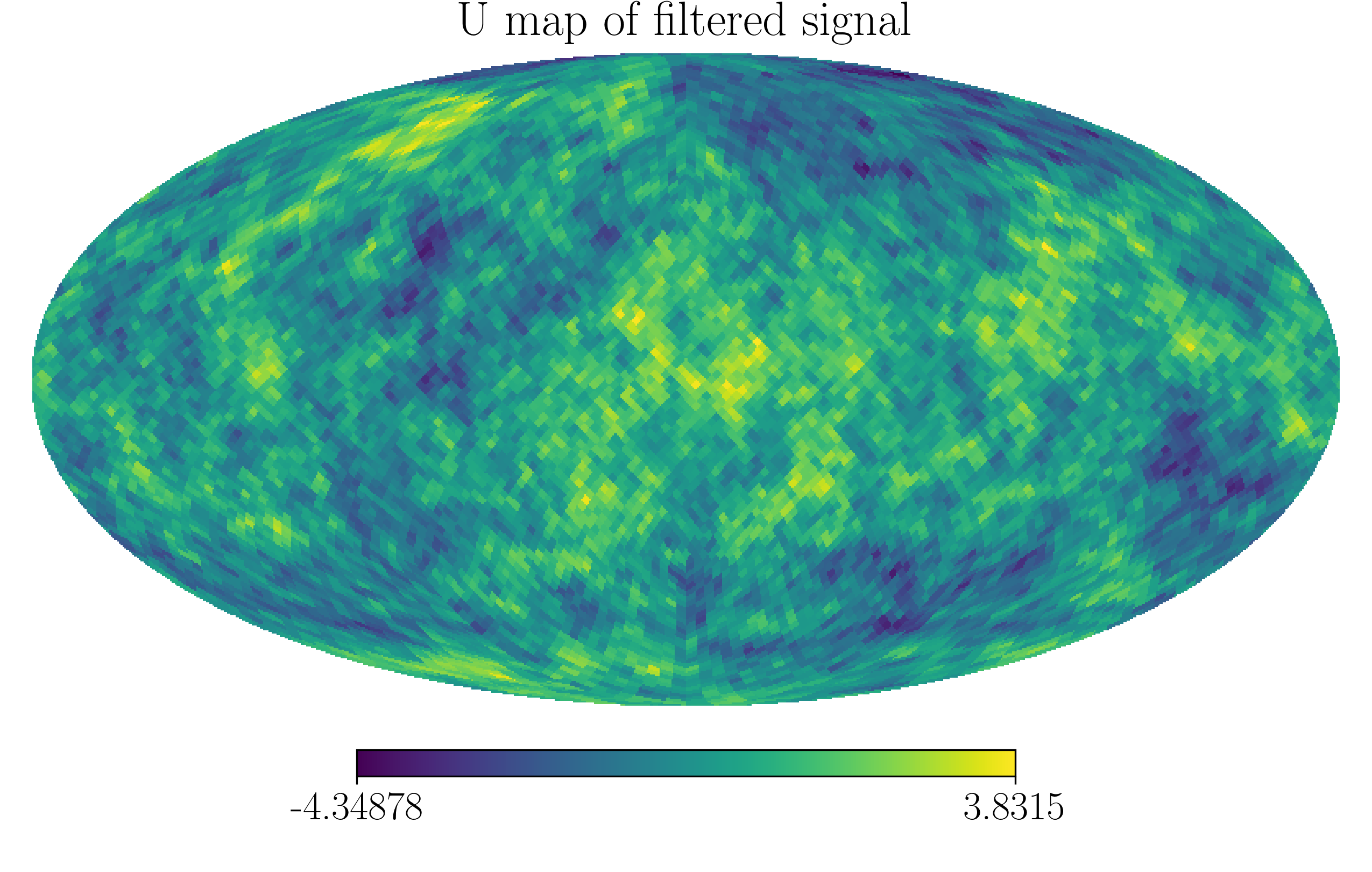}}
        \subfloat[Difference map $\hat{s}-\bar{s}$]{\includegraphics[width=0.33\textwidth]{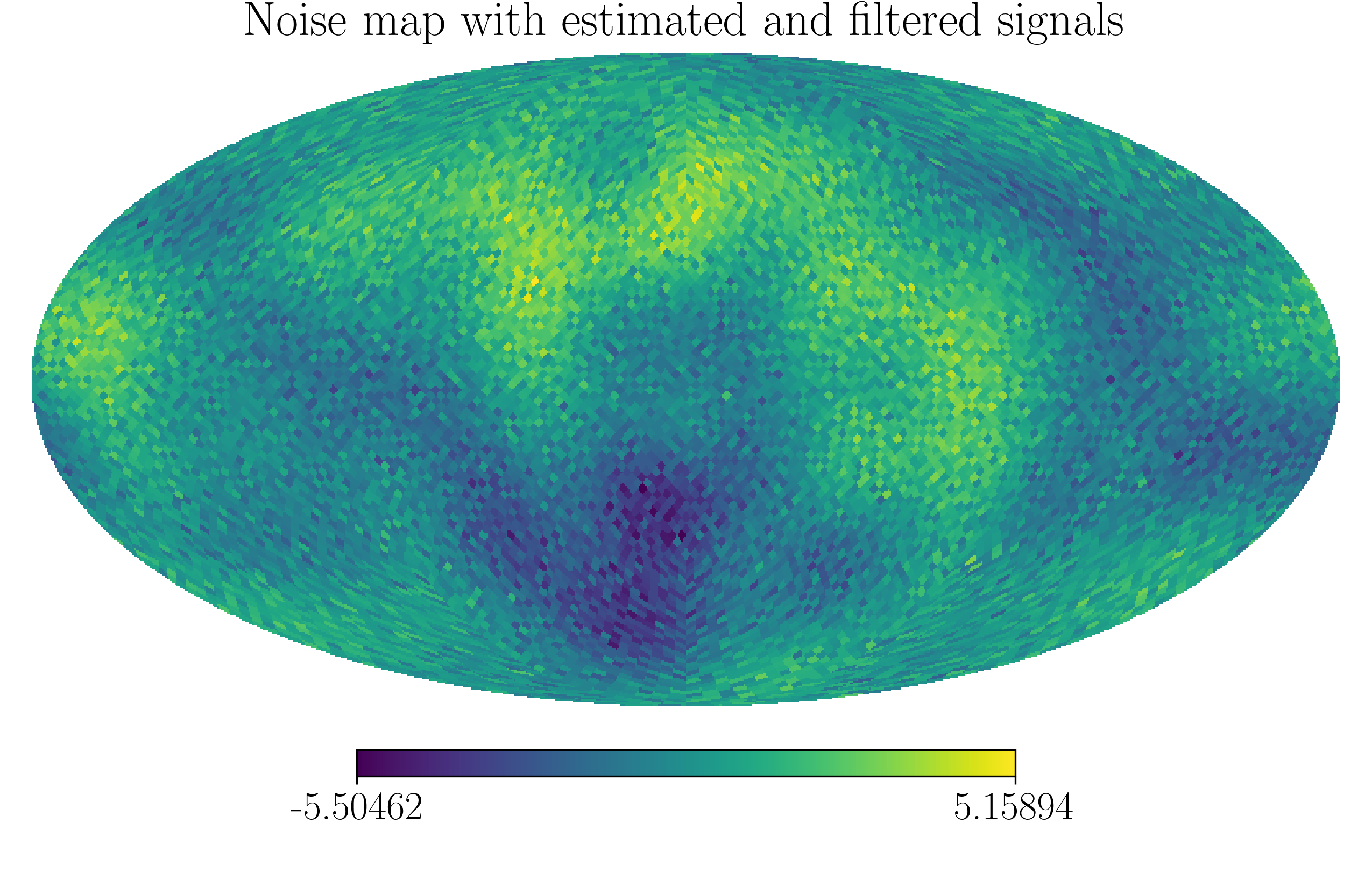}}
        \caption{\label{fig:GC}Second-best linear combination of wavelengths from a simulation using the Great Circle pattern. \change{The large scale structure in the difference map $\hat{s} - \bar{s}$ gives visual proof to the claim that this antenna pattern produces the least reliable reconstructions.}}
    \end{figure*}

    Fig. \ref{fig:1D} shows the second-best linear combination of wavelengths for the 1-Dimensional Interferometer pattern, as shown in Fig. \ref{fig:eigenvectors_1D}. Fig. \ref{fig:GC} shows the second-best linear combination of wavelengths for the Great Circle pattern, as shown in Fig. \ref{fig:eigenvectors_GC}. These maps give visual proof to the claim that the Great Circle antenna pattern produces the least reliable linear combinations of wavelengths.
    
    \subsection{Half Sky}
    The results of the previous section apply to an idealized isotropic all-sky experiment. In this section, we show results for a hypothetical experiment with partial sky coverage.
    As described in Section \ref{introduction}, we continue to perform the calculation of $\mathbf{\bar{M}}^{-1}$ in a spherical harmonic basis -- that is, the wavelength linear combinations we choose to reconstruct are optimized as if we were doing an all-sky isotropic experiment.
    
   For the examples illustrated here, we assume the observations cover half the sky: the pointings are chosen randomly and isotropically as before, but with pointing centers restricted to the northern hemisphere $\theta<\pi/2$.
   
    \begin{figure*}
        \centering
        \subfloat[Pure signal, $\vec{s}$]{\includegraphics[width=0.33\textwidth]{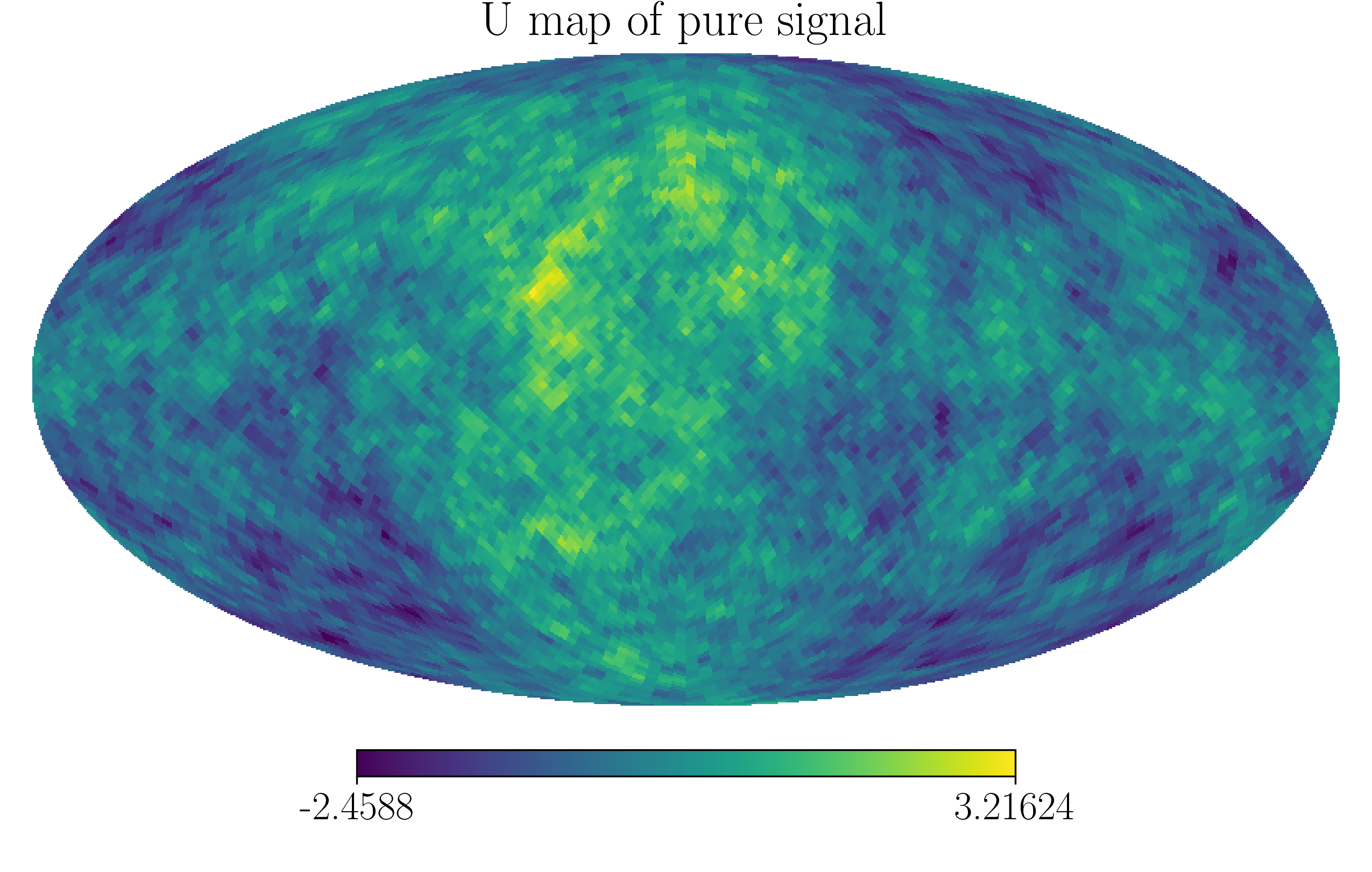}}
        \subfloat[Estimated signal, $\hat{s}$]{\includegraphics[width=0.33\textwidth]{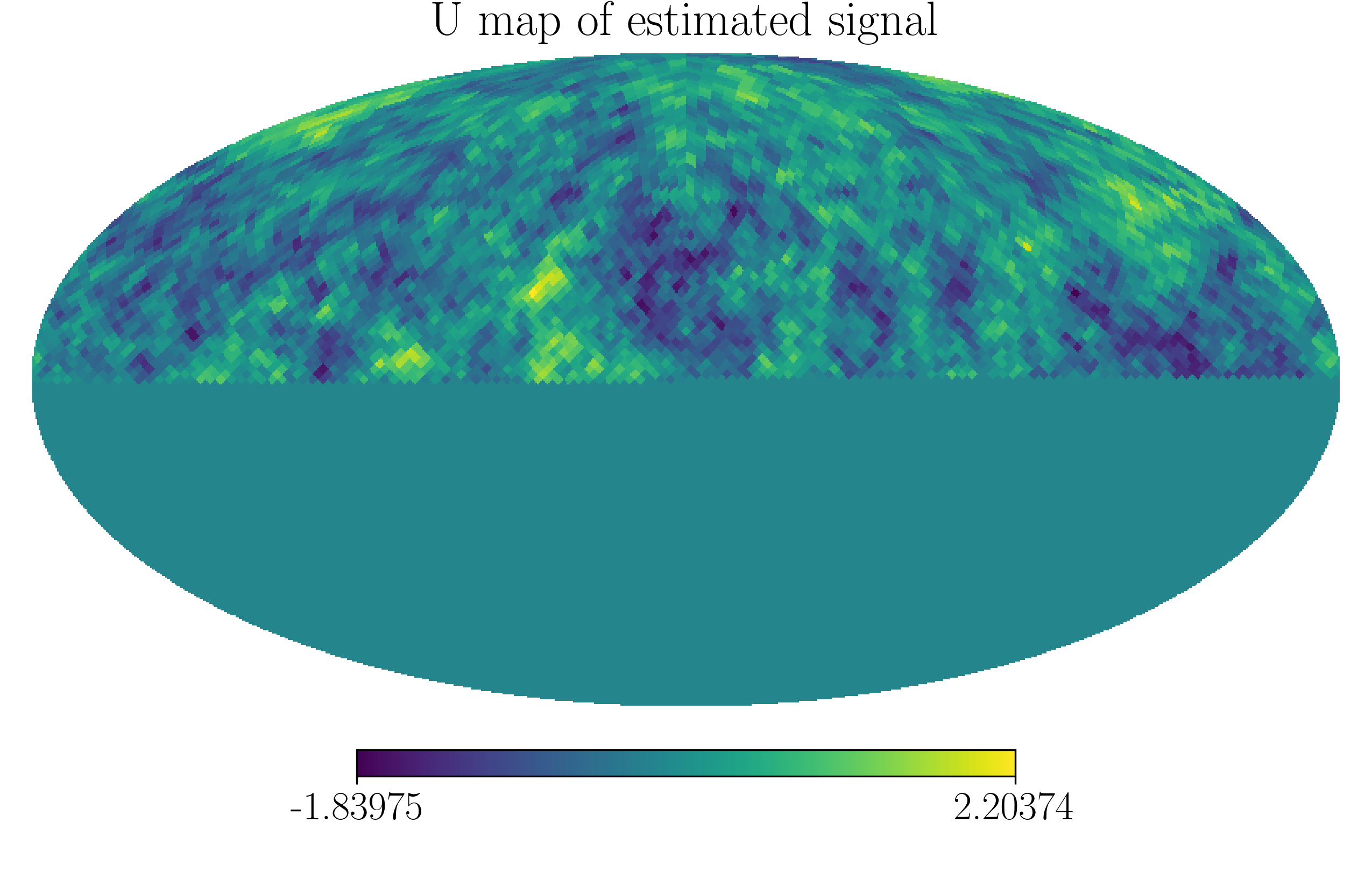}}
        \subfloat[Difference map $\hat{s} - \vec{s}$]{\includegraphics[width=0.33\textwidth]{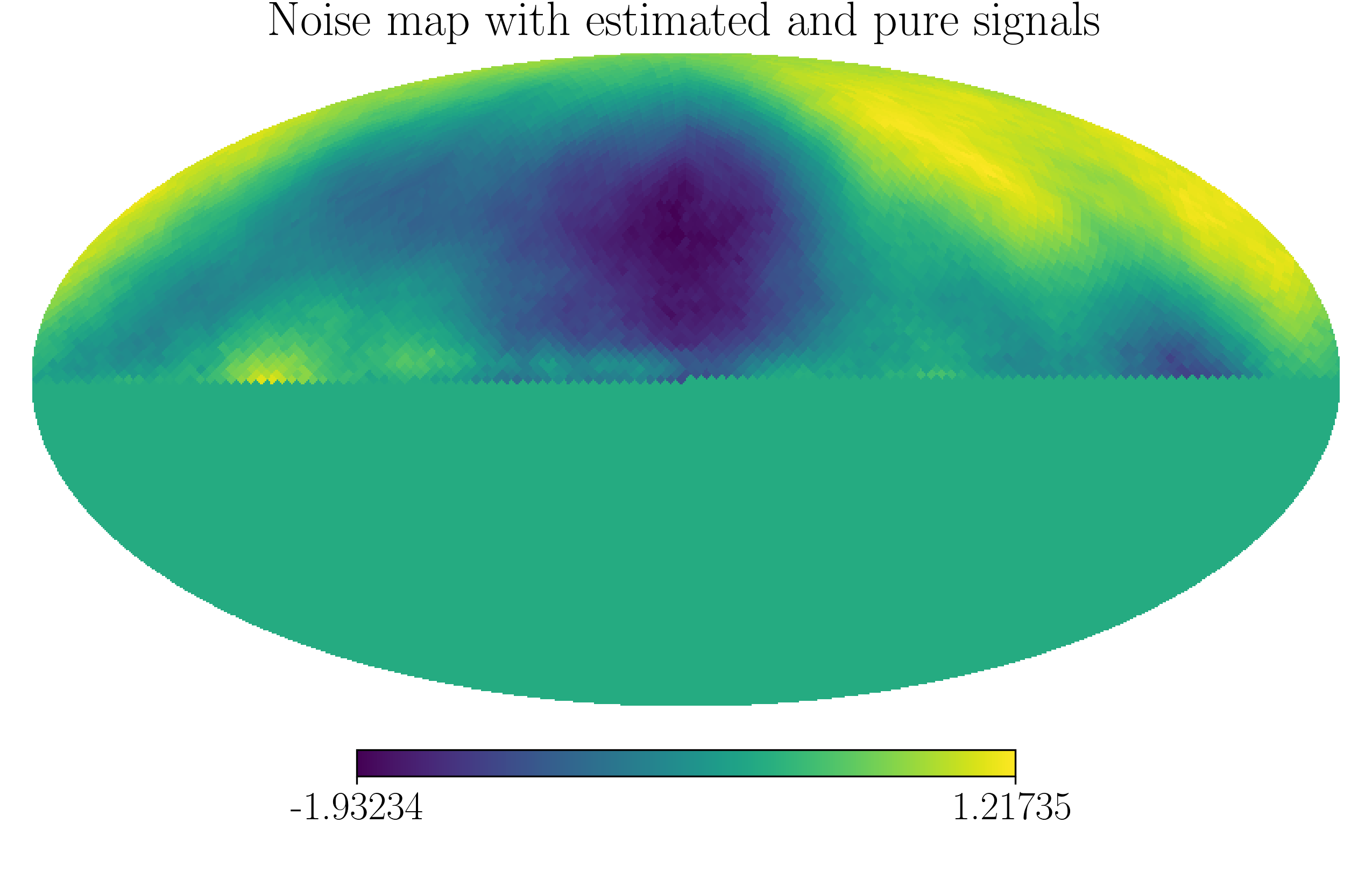}}

        \subfloat[Pure filtered signal, $\bar{s}$]{\includegraphics[width=0.33\textwidth]{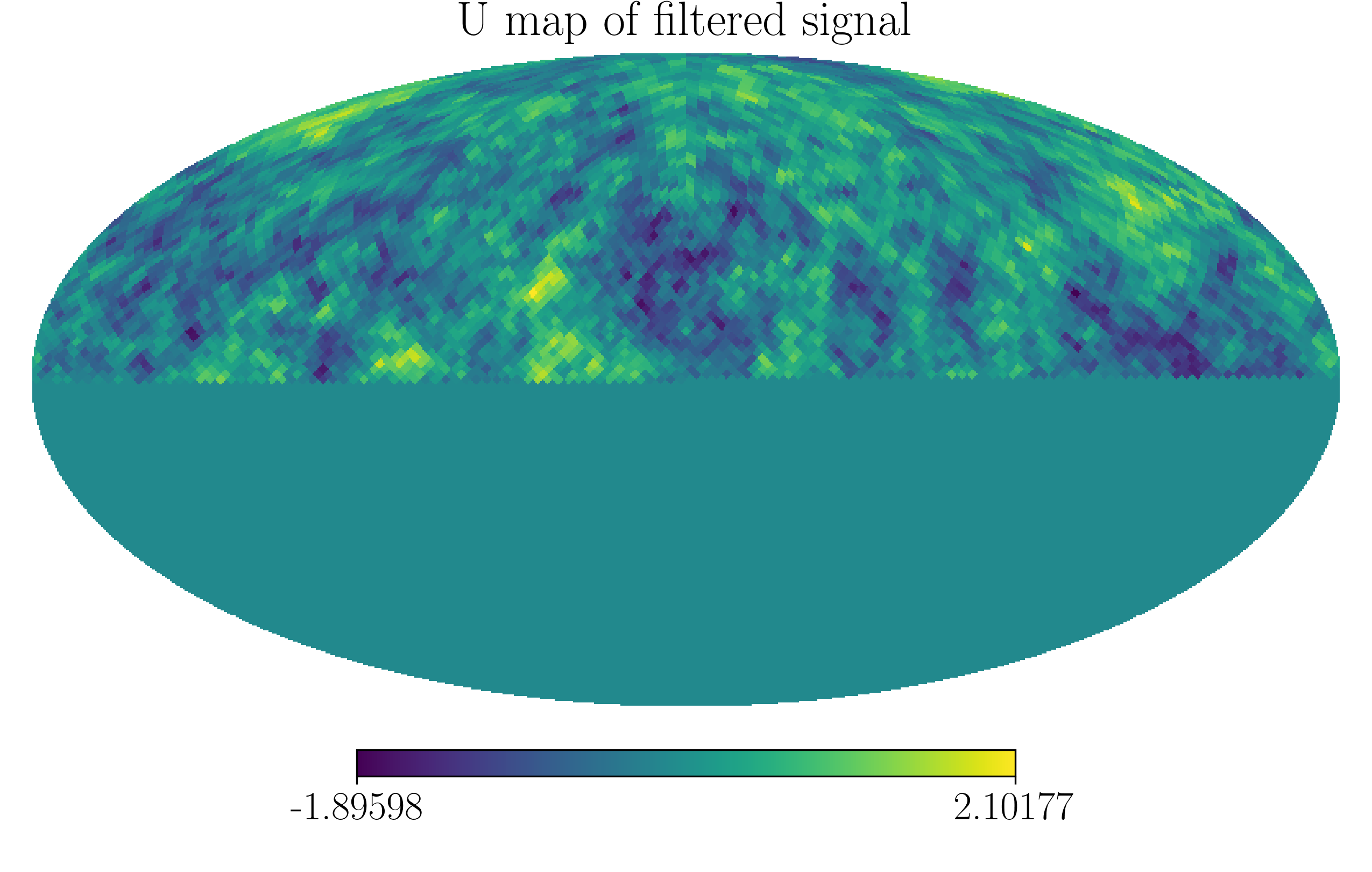}}
        \subfloat[Difference map $\hat{s} - \bar{s}$]{\includegraphics[width=0.33\textwidth]{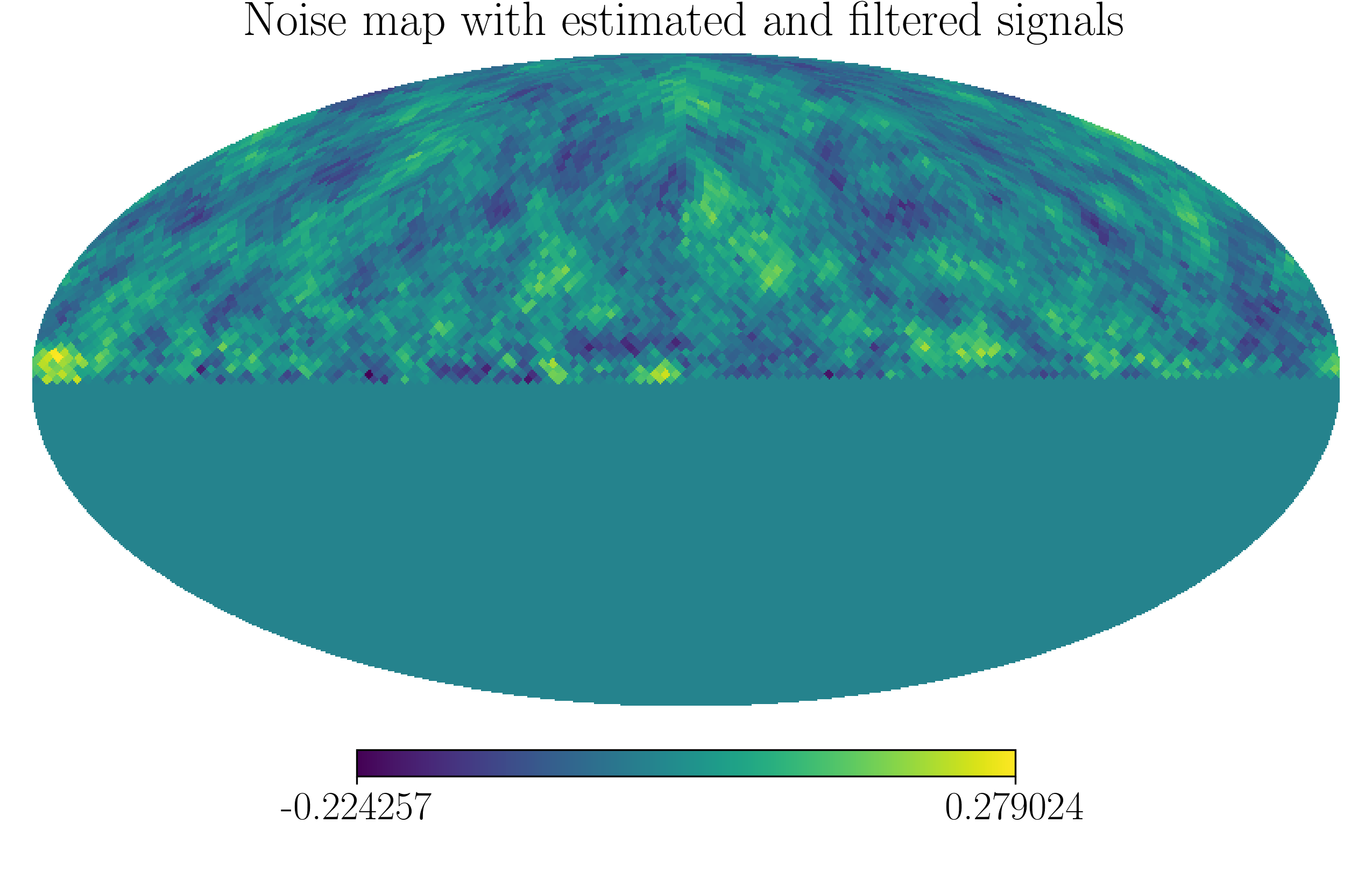}}
        \caption{\label{fig:2dhalf_sky}Second-best linear combination of wavelengths from a simulation over half the sky using the 2-Dimensional Interferometer pattern.}
    \end{figure*}
    Fig. \ref{fig:2dhalf_sky} shows a half-sky reconstruction for the 2-Dimensional Interferometer pattern.

    \begin{figure*}
        \centering
        \subfloat[Estimated signal, $\hat{s}$]{\includegraphics[width=0.33\textwidth]{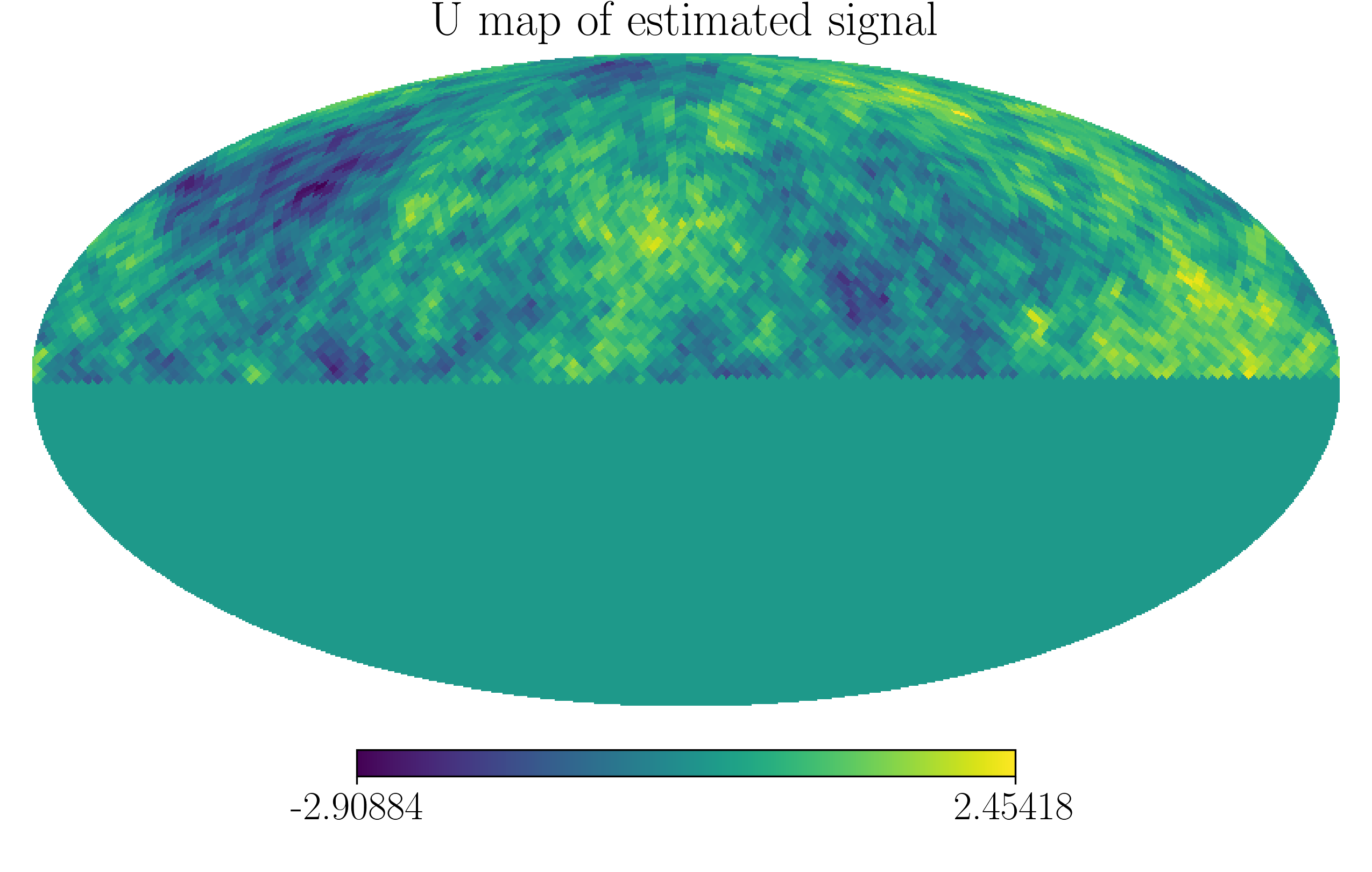}}
        \subfloat[Pure filtered signal, $\bar{s}$]{\includegraphics[width=0.33\textwidth]{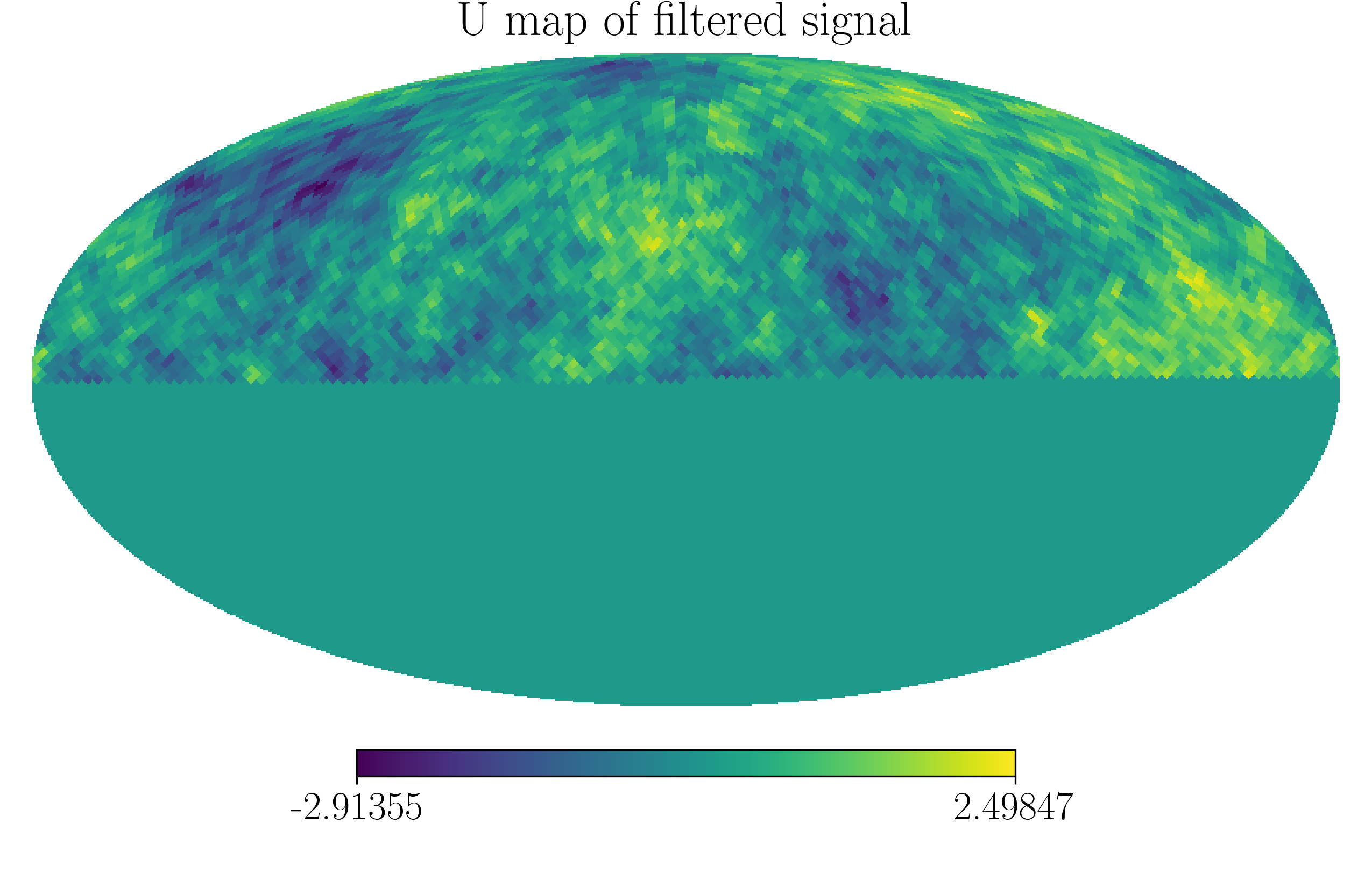}}
        \subfloat[Difference map $\hat{s} - \bar{s}$]{\includegraphics[width=0.33\textwidth]{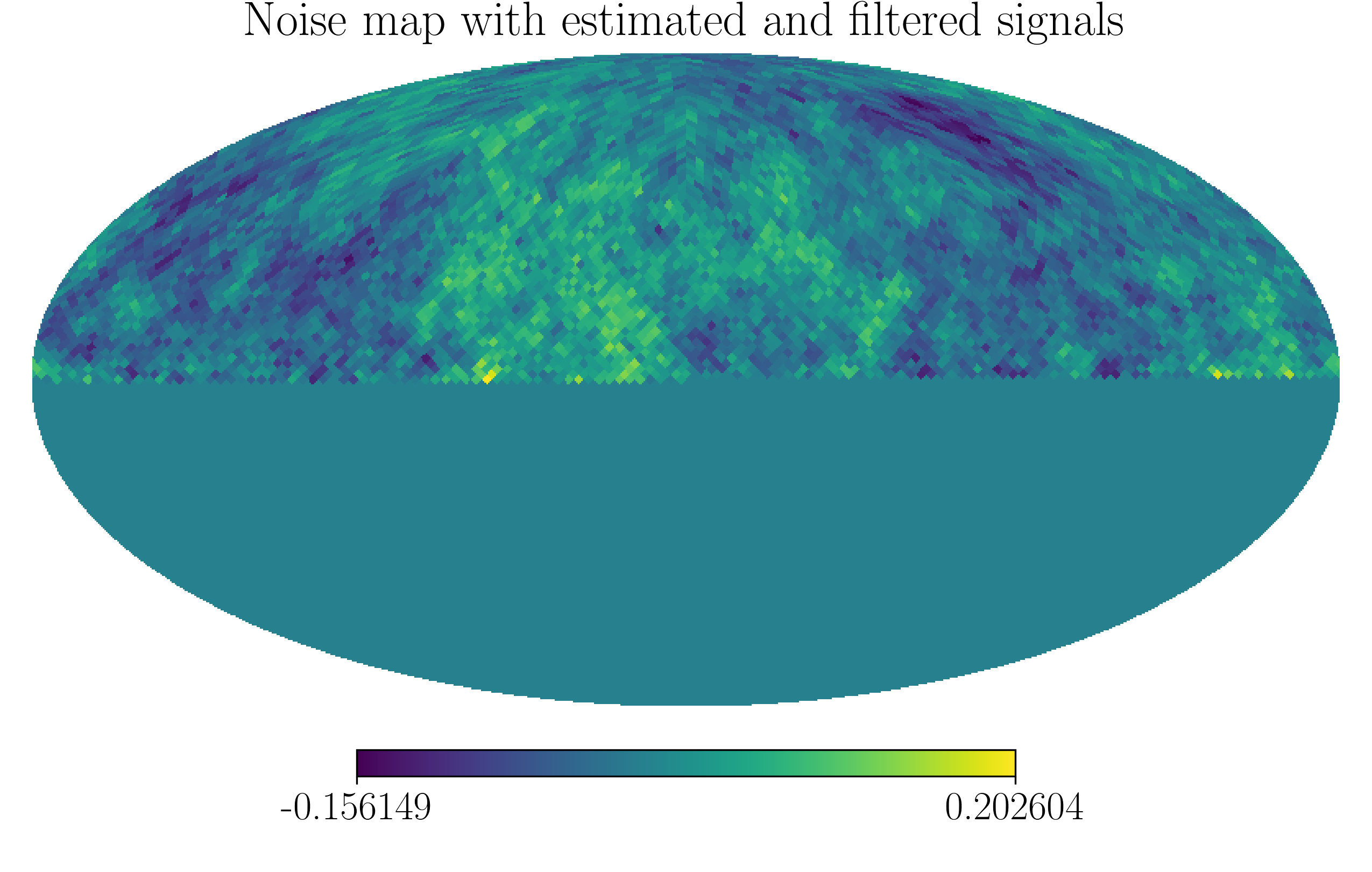}\label{fig:1dhalf_sky_dif}}
        \caption{\label{fig:1dhalf_sky}Second-best linear combination of wavelengths from a simulation over half the sky using the 1-Dimensional Interferometer pattern.}
    \end{figure*}
    
    The stark asymmetry of the 1-Dimensional and Great Circle antenna patterns compared to the 2-Dimensional antenna pattern makes it more difficult to achieve an isotropic experiment with fewer timesteps, namely the ability to scan the sky in all possible orientations of the telescope. Thus, for half-sky reconstructions, we must increase the number of timesteps for the 1-Dimensional and Great Circle antenna patterns. Fig. \ref{fig:1dhalf_sky} shows a half-sky reconstruction for the 1-Dimensional antenna pattern with five times more timesteps than in the full-sky reconstruction. While a qualitative look at these maps does not show large differences between the filtered and estimated signals, the noise map in Fig. \ref{fig:1dhalf_sky_dif} shows once again the largely random noise we expect between the two signals.
    \begin{figure*}
        \centering 
        
        \subfloat[Estimated signal, $\hat{s}$]{\includegraphics[width=0.33\textwidth]{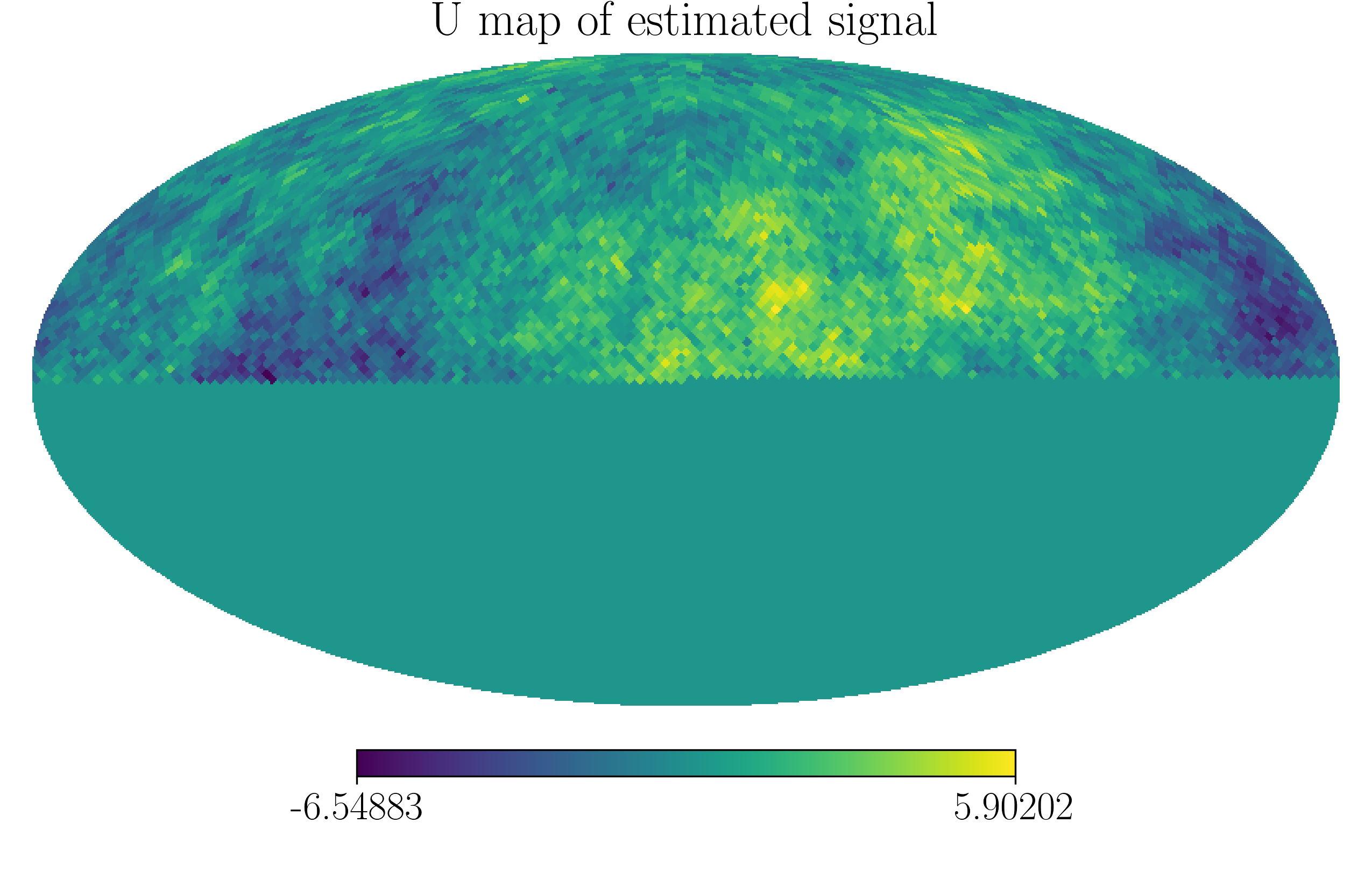}}
        \subfloat[Filtered signal, $\bar{s}$]{\includegraphics[width=0.33\textwidth]{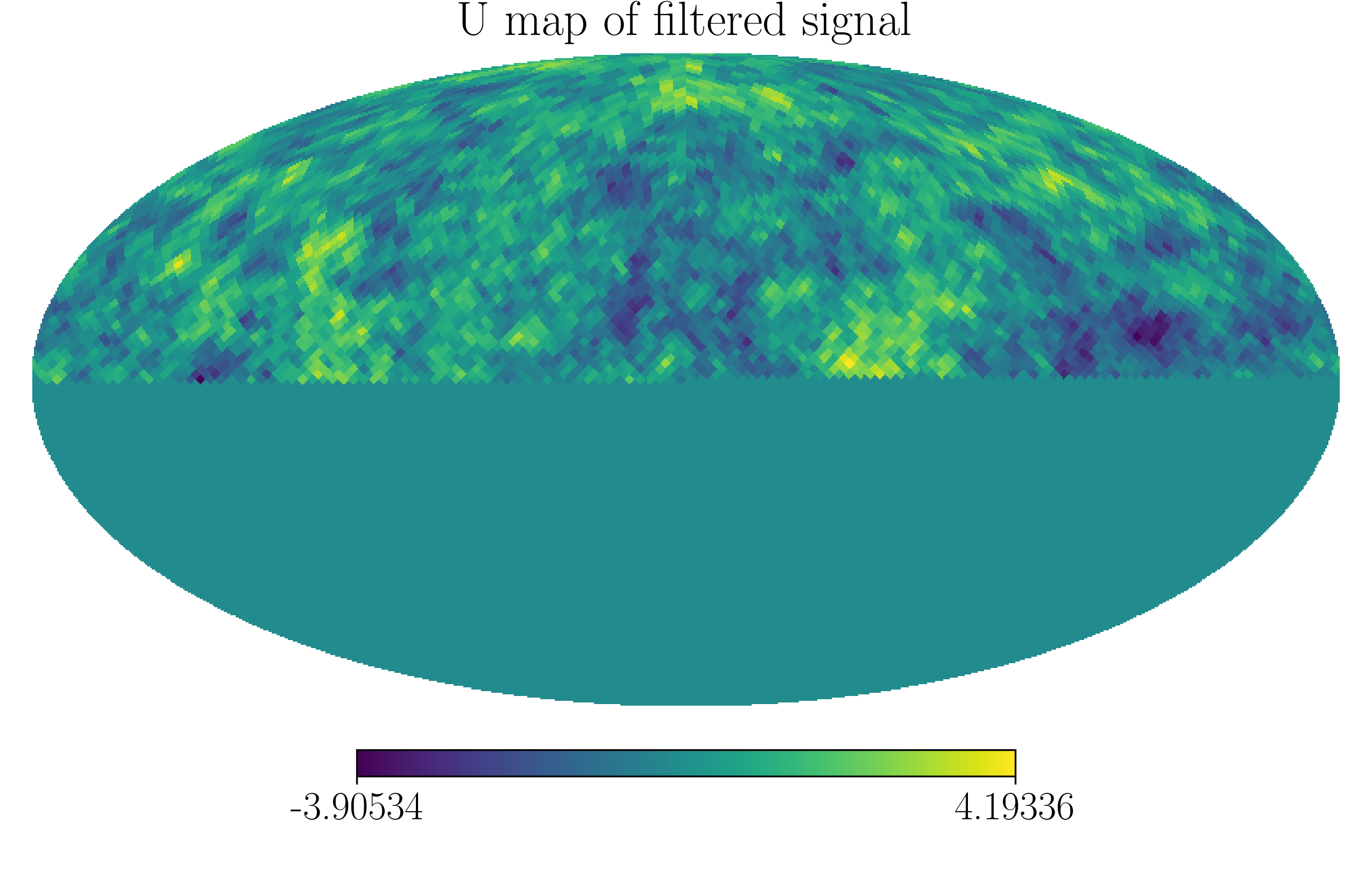}}
        \subfloat[Difference map $\hat{s} - \bar{s}$]{\includegraphics[width=0.33\textwidth]{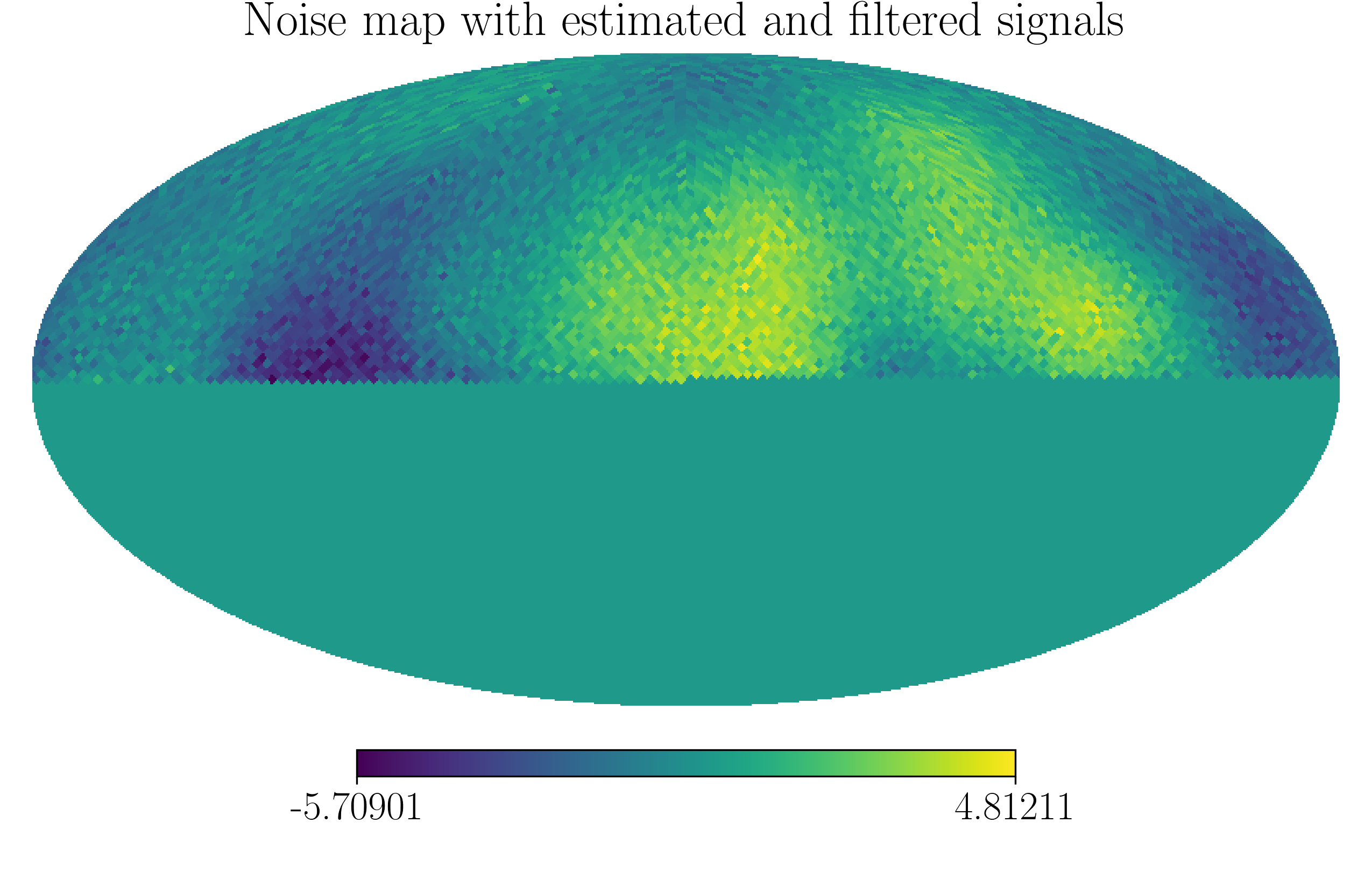}}
        \caption{\label{fig:gchalf_sky}Second-best linear combination of wavelengths from a simulation over half the sky using the Great Circle pattern.}
    \end{figure*}
    Fig. \ref{fig:gchalf_sky} shows a half-sky reconstruction for the Great Circle antenna pattern. For the purpose of time, this simulation was run with the same number of timesteps as in the full-sky case. If this antenna pattern were to be used for polychromatic reconstruction, a larger data set would be required to yield more accurate results.
    
    \subsection{Noise covariance}
    
    In this section we illustrate the noise properties of the reconstructed maps. The matrix $\mathbf{M}^{-1}$ is the noise covariance matrix for the entire multiwavelength signal. We can represent this matrix in the eigenbasis in wavelength space (while leaving it unaltered in pixel space). Then each row of the matrix will show the covariance of the noise in each reconstructed signal at a given pixel. Because $\mathbf{M}$ is sparse, any given row is easily found via the conjugate gradient method.
    
\begin{figure*}
        \centering
        \subfloat[Best linear combination]{\includegraphics[width=0.33\textwidth]{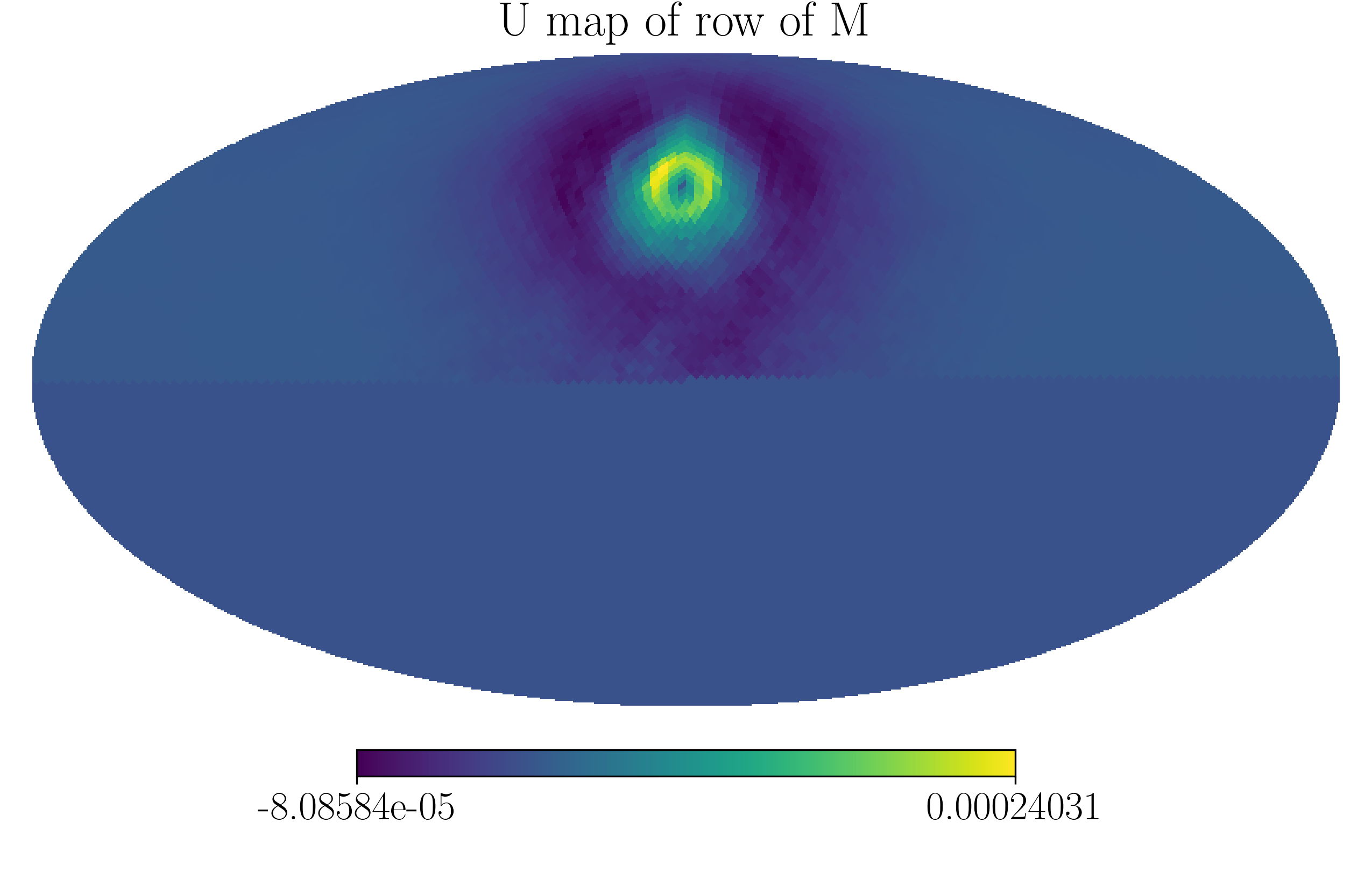}}
        \subfloat[Second best linear combination\label{fig:row_m_second}]{\includegraphics[width=0.33\textwidth]{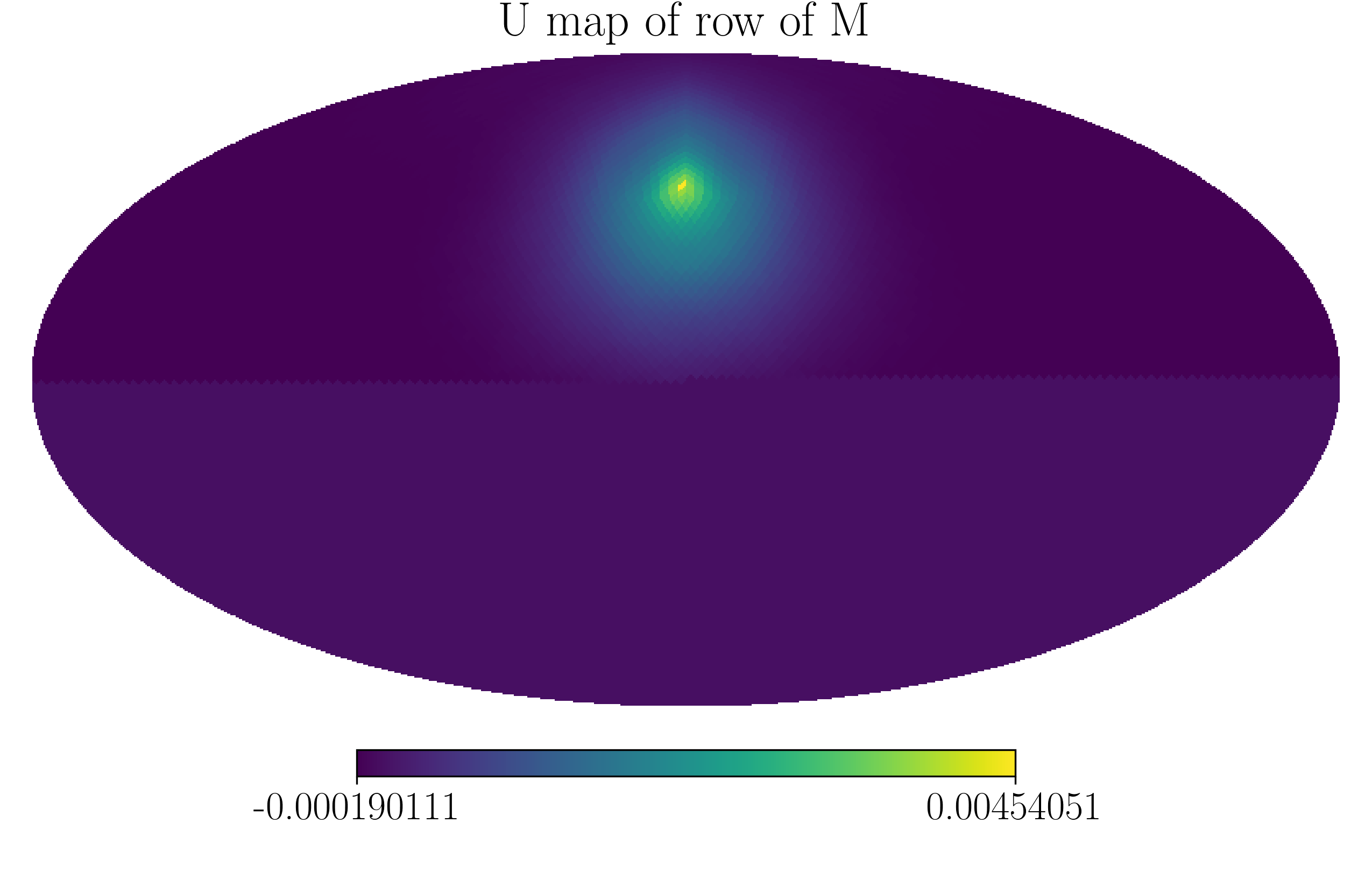}}
        \subfloat[Third best linear combination]{\includegraphics[width=0.33\textwidth]{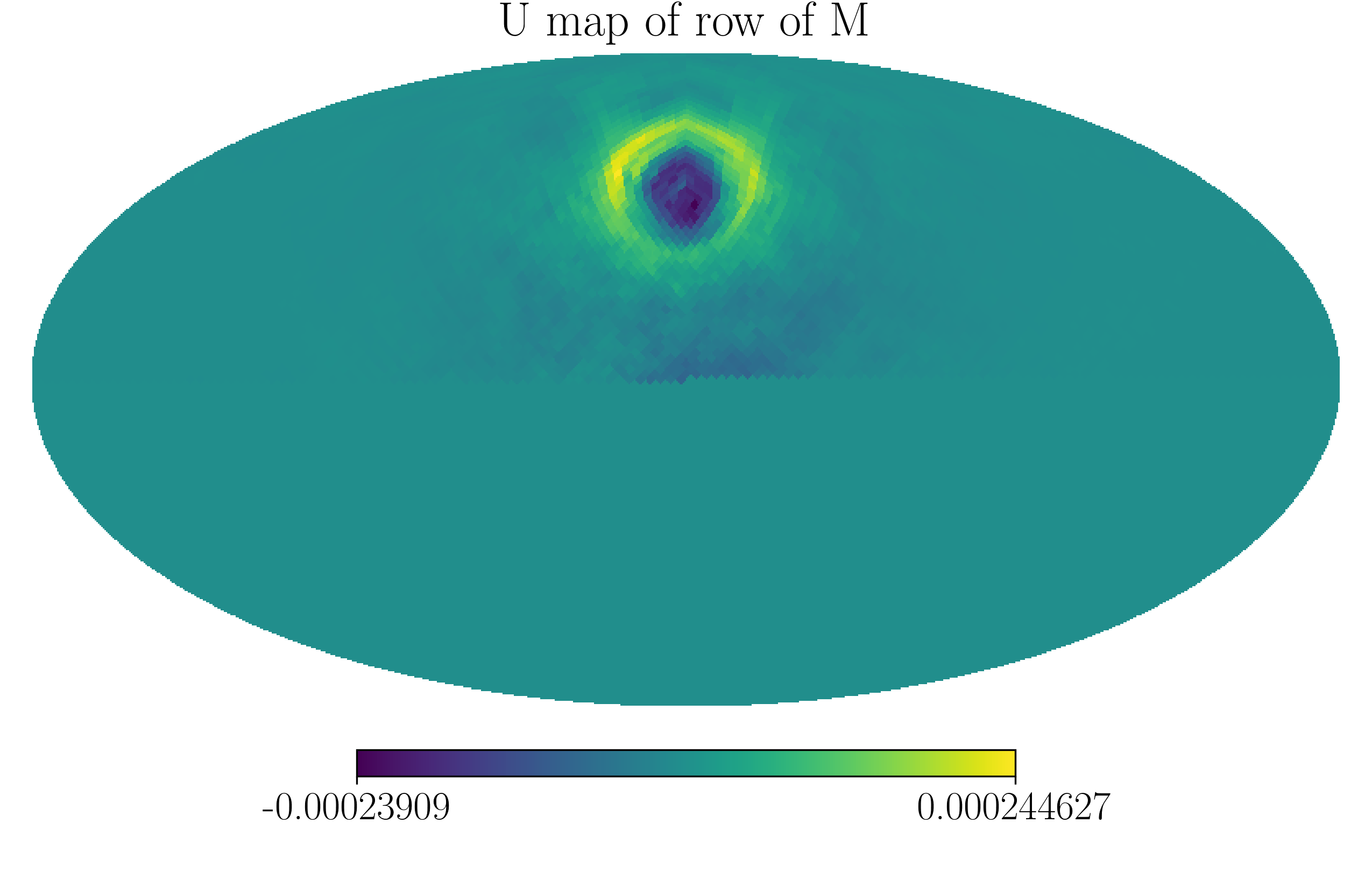}}
        \caption{\label{fig:m_row}One row of \change{the noise covariance matrix} $\mathbf{M}$ in a simulation over half the sky using the 2-Dimensional Interferometer antenna pattern. \change{This row corresponds}  to a pixel in the northern hemisphere and to the second-best wavelength combination  \change{The middle panel shows the noise has some spatial correlation, and the first and third panels show there is some nonzero cross-correlation between reconstructed wavelength combinations.}}
    \end{figure*}

    Figure \ref{fig:m_row} illustrates the result of such a calculation. The row of the noise covariance matrix corresponds to a particular pixel in the northern hemisphere and to the second-best wavelength combination. The maximum value in the middle panel is thus the noise variance for this reconstructed signal value. The structure in the middle panel shows that the noise has some spatial correlation, and the nonzero values in the first and third panel show that there are nonzero but weak cross-correlations between the different reconstructed wavelength combinations.

\begin{figure}
        \centering
        \includegraphics[width=0.4\textwidth]{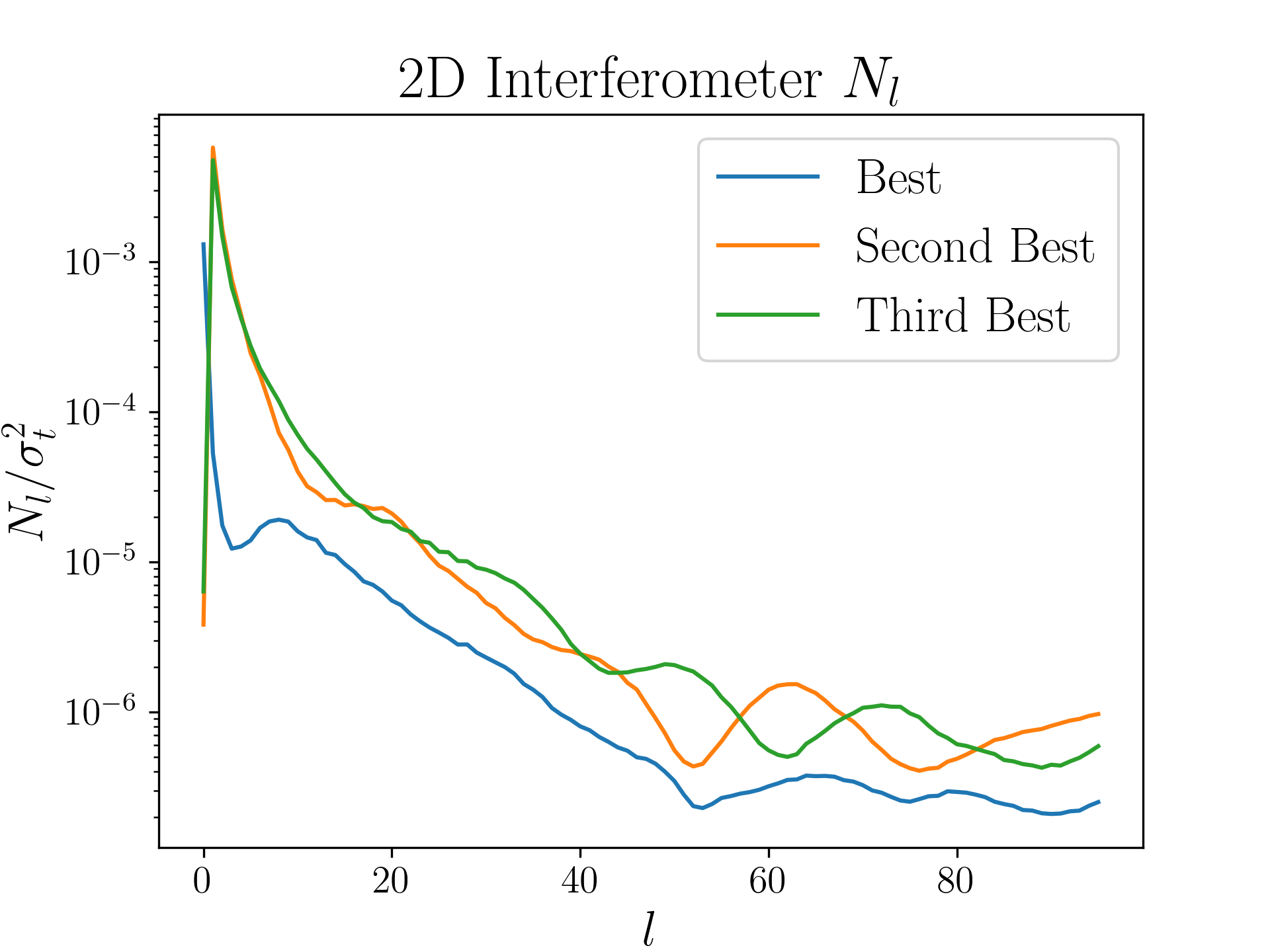}
        \caption{\change{Noise power of the three best linear combinations of wavebands from a simulation using the 2-Dimensional Interferometer pattern. This data was obtained by averaging the power spectra of the reconstructed maps in 100 simulations made using a data vector $\vec{d}$ consisting of white noise.}}
        \label{fig:noise_power}
    \end{figure}

        \change{
        Another way to visualize the spatial structure of the noise is via the noise power spectra $N_l$ for the various reconstructed maps. 
   Fig. \ref{fig:noise_power} shows $N_l$ for the three best linear combinations of wavebands using the 2-Dimensional Interferometer pattern. These were computed by applying the polychromatic map reconstruction algorithm to 100 sets of white-noise TOD. The noise power spectrum $N_l$ for each reconstructed color is the average of the power spectra of the reconstructed maps.
    }
    
\section{Conclusion}
\label{sec:conclusion}
   
We have presented a formalism for finding the optimal ``colors'' that can be extracted from a total-intensity broadband telescope that scans the sky with an asymmetric beam pattern. This approach is inspired by CMB observations, which often involve broadband detectors covering a small number of wavelength bands. In analyzing CMB sky maps, color information is important for component separation but is unfortunately often in short supply. The ability to squeeze extra color information out of such maps is therefore of particular importance. 

The underlying idea behind this method was developed as part of the design of the QUBIC telescope, and experiments of this sort, whose antenna patterns are highly asymmetric, are most likely to benefit from it.

Although the method is inspired by CMB studies, color information is of course invaluable in many areas of astrophysics, so it is quite possible that the methods developed herein will have broader applications.

We have shown that for some antenna patterns, color maps can be produced that have noise levels of the same order of magnitude as the total-intensity maps, indicating that significant additional information is likely to be available.

The colors that we find via this method are optimal for the specific case of an all-sky isotropic scan pattern. Real scans do not necessarily achieve this coverage. However, we have demonstrated in our simulations that the noise levels achieved in partial-sky experiments are similar to those in the all-sky case, as expected.

\change{Polychromatic mapmaking is possible in practice for observations that scan the sky with broadband detectors, using an instrument with an asymmetric beam. CMB observations are the most natural arena in which to apply this method: CMB detectors are generally broadband; color information is vital for foreground separation, and some CMB instruments have asymmetric beams. The QUBIC experiment in particular is the inspiration for this work: one of the authors is a member of the QUBIC collaboration, and this work builds on methods developed for QUBIC \cite{stolpovskiy,qubic2}. We intend to explore implementation of the methods developed here into the QUBIC analysis pipeline.}

\change{Although QUBIC is the main use case we envision, the method may be applicable in other contexts. The highly asymmetric antenna patterns that make the method most promising are likely to be found in other interferometric contexts or possibly for experiments such as CHIME \cite{chime} with long narrow antennas.}

Having shown that this method is promising, we envision a number of ways in which it can be extended:
\begin{itemize}
    \item The method can be tested on simulations of other antenna patterns and scan strategies, including, for example, a more realistic model of QUBIC.
    \item The method can be applied to maps containing realistic astrophysical components -- e.g., a simulation including CMB fluctuations and foreground contaminants such as dust and synchrotron.
    \item Generalization of the method to polarization data should be straightforward. The foreground-removal problem is particularly acute for CMB polarization experiments, as the polarization signal is weak and the polarization properties of the foregrounds are more uncertain (e.g., \cite{vonhausegger}).
\end{itemize}

\section*{Acknowledgments}

Early stages of this work were supported by NSF Award 141013 and by a visitor grant from the Institut Lagrange (Paris). Subsequent work was supported by multiple Research Fellowships from the University of Richmond School of Arts \& Sciences. We acknowledge enlightening conversations with M. Stolpovskiy, J.-Ch. Hamilton, R. Charlassier, and J. Kaplan.

\bibliography{references}

\appendix


    \section{\label{eigenvector_proof}Eigenvalues and Eigenvectors of Covariance Matrices} 
        Let $\hat s$ be an unbiased estimator of some signal vector $\vec s$, with uncertainties characterized by a covariance matrix $\mathbf{M}$ with entries
          \begin{equation}
            M_{dd'}=\langle\delta\hat{s}_d\delta\hat{s}_{d'}\rangle
            =\langle\delta\hat{s}\delta\hat{s}^T\rangle_{dd'}.
        \end{equation}
    
Suppose that we want to estimate $u=\vec v\cdot\vec s$ for some vector $\vec{v}$. We define an estimator
        \begin{equation}
            \hat u=\sum_{d}v_d \hat{s}_d.
        \end{equation}
        Then
\begin{equation}
            \begin{split}
            \langle(\delta\hat{u})^2\rangle & = \langle(\vec{v}\cdot\delta\hat{s})^2\rangle \\
            & = \langle(\vec{v}^T\delta\hat{s})(\delta\hat{s}^T\vec{v})\rangle \\
            &=\vec{v}^T\langle\delta\hat{s}\delta\hat{s}^T\rangle\vec{v} \\
            &=\vec{v}^T\mathbf{M}\vec{v}.\label{eq:duvmv}
            \end{split}
\end{equation}
        We want to choose the ``best'' vector $\vec v$ -- that is, the one that minimizes the uncertainty -- 
        subject to the constraint $|\vec{v}|=1$, or $\sum_d\vec{v}_d^{\,2}=1$, for
        \begin{equation}
            \delta\hat{u}^2  = \sum_{d,g}{v}_dM_{dg}{v}_g.
        \end{equation}
        We differentiate with respect to an arbitrary component $v_h$ such that
        \begin{equation}
            \frac{\partial(\delta\hat{u}^2)}{\partial {v}_h}  = \sum_gM_{hg}{v}_g + \sum_d{v}_dM_{dh} 
            =2(\mathbf{M} {v})_h.
        \end{equation}
        Introducing a Lagrange multiplier to apply the constraint gives
        \begin{equation}
            (\mathbf{M}{v})_h = \lambda {v}_h.
        \end{equation}
The optimal linear combination of data points is thus an eigenvector of the covariance matrix. From equation (\ref{eq:duvmv}), it follows that
        \begin{equation}
            \delta\hat{u}^2=\vec{v}^T\lambda\vec{v}
            =\lambda\vec{v}^T\vec{v} =\lambda.
        \end{equation}
        The eigenvalue of the covariance matrix is thus the uncertainty variance in the linear combination corresponding to the eigenvector. The optimal linear combination is the eigenvalue corresponding to the smallest eigenvector.


    \section{\label{sec:c_list}Covariance Matrices in Spherical Harmonic Space} 
        Assuming white noise ($\mathbf{N}=\sigma_t^2\mathbf{1}$), an arbitrary element of the harmonic-space inverse noise covariance matrix $\tilde{\mathbf{M}}^{-1}$ can be written
        \begin{equation}
            \tilde{M}^{-1}_{(lmf)(l'm'f')}=
            \sigma^{-2}_t\sum_t\left(\sum_p\mathcal{A}(\mathbf{R}_t(\hat{r}_p);\lambda_f)A_{\mathrm{pix}}Y_{lm}(\hat{r}_p)
            \sum_{p'}\mathcal{A}(\mathbf{R}_t(\hat{r}_{p'});\lambda_{f'})A_{\mathrm{pix}}Y_{l'm'}(\hat{r}_{p'})^*\right).
        \end{equation}
        We assume that the pixelization is fine enough that sums over pixels can be converted to integrals. Moreover, we assume an isotropic experiment, meaning that the rotation matrices $\mathbf{R}_t$ densely and uniformly sample SO(3), the space of all possible rotation matrices. Under this assumption, we can also convert the sum over $t$ to an integral. The result is
\begin{equation}\label{eq:mtinv}
            \tilde{M}^{-1}_{(lmf)(l'm'f')} = \frac{N_t}{8\pi^2\sigma_t^2}\int_{SO(3)} d\mathbf{R}\, f_{lmf}(\mathbf{R})f_{l'm'f'}^*(\mathbf{R}),
\end{equation}
        where $N_t$ is the number of timesteps, $8\pi^2$ is the volume of the space SO(3), and
\begin{equation} \label{eq:appendixb1}
            f_{lmf}(\mathbf{R}) = \int d^2\vec{r} \,\mathcal{A}(\mathbf{R}(\vec{r}); \lambda_f)Y_{lm}(\vec{r}).
\end{equation} 
        Expanding the antenna pattern in spherical harmonic coefficients gives
\begin{equation}
            \mathcal{A}(\mathbf{R}(\vec{r});\lambda_f) = \sum_{L,M} A_{LMf} Y_{LM}(\mathbf{R}(\vec{r})).    
\end{equation}
        Substituting this into equation \ref{eq:appendixb1} gives
\begin{equation} \label{eq:appendixb2}
            f_{lmf}(\mathbf{R}) = \sum_{L,M} A_{LMf} \int Y_{LM}(\mathbf{R}(\vec{r}))Y_{lm}(\vec{r})d^2\vec{r}.
\end{equation}
        The rotation of the spherical harmonic matrix $Y_{lm}$ can be expressed in terms of a Wigner matrix, 
\begin{equation}
            Y_{lm}(\mathbf{R}(\vec{r})) = \sum_{m'} D^l_{mm'}(\mathbf{R})Y_{lm'}(\vec{r}).
\end{equation}
        Substituting this into equation \ref{eq:appendixb2} gives
\begin{equation}
        f_{lmf}(\mathbf{R}) = \sum_{L,M} A_{LMf} D^L_{MM'}(\mathbf{R}) \int Y_{LM'}(\vec{r})Y_{lm}(\vec{r}) d^2\vec{r}
        =\sum_MA_{lMf}D^l_{Mm},
\end{equation}
using the orthonormality of the spherical harmonics.

        Substituting this expression into equation (\ref{eq:mtinv}) gives
        \begin{equation}
       \tilde{M}^{-1}_{(lmf)(l'm'f')} =
       \frac{N_t}{8\pi^2\sigma_t^2}\sum_{M,M'} A_{lMf}A^*_{l'M'f'}\int_{\mathrm{SO(3)}} d\mathbf{R}
       D^l_{Mm}(D^{l'}_{M'm'})^*.
       \end{equation}
        This integral is $(8\pi^2/(2l+1))\delta_{ll'}\delta_{MM'}\delta_{mm'}$ \cite{zare}, leading to the conclusion that
\begin{equation}
        M^{-1}_{(lmf)(l'm'f)'} = \delta_{ll'}\delta_{mm'}
        \frac{N_t}{\sigma^2_t} \frac{1}{2l+1} \sum_{M} A_{lMf}A_{lMf'}^*.
\end{equation}
        The matrix $\tilde{\mathbf{M}}^{-1}$ is thus block diagonal in $lm$, with blocks that depend only on $l$, not $m$.

\end{document}